\documentclass[rmp,aps,nofootinbib,endfloats,preprint]{revtex4}
\usepackage{graphicx}
\usepackage{dcolumn}
\usepackage{bm}
\usepackage{amsfonts}

\newcommand{\bi}{\begin{itemize}}
\newcommand{\ei}{\end{itemize}}

\newcommand{\bea}{\begin{eqnarray}}
\newcommand{\eea}{\end{eqnarray}}
\newcommand{\be}{\begin{equation}}
\newcommand{\ee}{\end{equation}}

\newcommand{\fd}{\dot{\varphi}}

\newcommand{\fdd}{\ddot{\varphi}}
\newcommand{\ld}{\dot{\lambda}}
\newcommand{\ldd}{\ddot{\lambda}}
\newcommand{\nd}{\dot{\nu}}
\newcommand{\ndd}{\ddot{\nu}}

\newcommand{\sdrs}{\alpha^{\prime}}
\newcommand{\sdr}{\sqrt{\alpha^\prime}}

\begin{document}

\title{String Gas Cosmology}
\author{Thorsten Battefeld}
\email{Battefeld@physics.brown.edu} \affiliation{Physics Department, Brown
University, Providence, RI 02912 USA.}
\author{Scott Watson}
\email{watsongs@physics.utoronto.ca} \affiliation{Physics Department,
University of Toronto, Toronto, ON, Canada M5S 1A7.}
\date{\today}

\begin{abstract}
We present a critical review and summary of String Gas Cosmology. We include a
pedagogical derivation of the effective action starting from string theory,
emphasizing the necessary approximations that must be invoked. Working in the
effective theory, we demonstrate that at late-times it is not possible to
stabilize the extra dimensions by a gas of massive string winding modes. We
then consider additional string gases that contain so-called enhanced symmetry
states. These string gases are very heavy initially, but drive the moduli to
locations that minimize the energy and pressure of the gas. We consider both
classical and quantum gas dynamics, where in the former the validity of the
theory is questionable and some fine-tuning is required, but in the latter we
find a consistent and promising stabilization mechanism that is valid at
late-times. In addition, we find that string gases provide a framework to
explore dark matter, presenting alternatives to $\Lambda$CDM as recently
considered by Gubser and Peebles. We also discuss quantum trapping with string
gases as a method for including dynamics on the string landscape.

\end{abstract}
\maketitle \tableofcontents

\section{Introduction}

String theory continues to have a number of challenges to address if it is to
be made experimentally verifiable. Cosmology offers an exciting opportunity to
explore such challenges, since the early universe provides conditions where
string dynamics would play a vital role.  To investigate the predictions of
string cosmology it is important to have concrete constructions of string
models in backgrounds that are compatible with our understanding of the early
universe.  In particular, this presents us with the challenge of finding
solutions of string theory in time-dependent backgrounds and at nonzero
temperature.

The usual method for constructing models of string cosmology is to compactify
any extra dimensions and then focus on the low energy, massless degrees of
freedom. However, this presents a problem since the low energy equations of
motion lack potentials to fix the massless moduli.  For cosmology, this implies
the existence of many light scalars, which if not fixed at late-times would
seem to contradict current observations. Nevertheless, a few light scalars
could prove valuable to cosmology, since they could address the issue of dark
energy, dark matter, or provide a theoretical motivation for inflation.

String or Brane Gas Cosmology (SGC) is an approach to string cosmology which
began with the pioneering work of Brandenberger and Vafa in
\cite{Brandenberger:1988aj}. They presented an elegant explanation for the
dimensionality of space-time by considering the effects of massive string modes
on the evolution of the early universe.  Since this seminal paper, considerable
effort has gone into realizing whether such a scenario is possible. In fact,
the cosmology of string gases has lead to interesting conclusions beyond those
originally proposed by Brandenberger and Vafa. In this paper we attempt to
present a pedagogical, yet critical, review of the string gas approach.

In Section II, we review the origin of the effective action of string cosmology
as it arises from string theory in the low energy - weak coupling limit. For
homogeneous fields, this effective theory exhibits a dynamical symmetry,
so-called scale factor duality.  We present the BPS fundamental string solution
and corresponding stress-energy tensor for the special case of a
time-independent background. In Section III, after explicitly stating the
assumptions of the SGC approach, we generalize the fundamental strings to a
time-dependent background treating them as an ideal gas.  We derive the
corresponding energy and pressure and discuss the duality properties of the
spectrum.  In Section IV, we return to the Brandenberger and Vafa mechanism and
review recent work that challenges the heuristic argument.  However, we point
out that this argument is not quintessential to string gas cosmology. Next, we
consider the effect of a classical string gas on the time-dependent background.
This has been examined in the literature from both the $10D$ String frame and
$4D$ Einstein frame perspectives. We review these works, stressing the
importance that physical quantities are frame independent. Using this, we
demonstrate that string gases of purely winding modes are not enough to
stabilize the extra dimensions.

A possible resolution to these problems consists of considering string states
that become massless at critical values of the radion (scale of the extra
dimension). These gases can drive the evolution of the radion to the location
which minimizes the pressure of the gas.  However, we will see that this
approach suffers from fine-tuning issues: first, each string gas configuration
can lead to a different attractor point on the moduli space; second, if the
radion starts far from the attractor point, the density of the gas will exceed
the energy cutoff of the effective theory, questioning the validity of the
approach.

In Section V, we present a resolution to these fine-tuning problems, by
considering the quantum aspects of the string gas.  This approach, known as
quantum moduli trapping \cite{Kofman:2004yc,Watson:2004aq}, takes the initial
theory to contain only the low energy massless modes of the string.  Then, as
the radion nears a point of enhanced symmetry (ESP) where additional states
become light, the states must be included in the effective action.  This leads
to particle production of the additional light states.  Once on-shell, these
states result in a confining potential, since their energy density grows as the
radion departs from the ESP.

In Section VI, we consider the possibility of obtaining observational
signatures from string gases. We demonstrate that remnant strings in the extra
dimensions provide natural candidates for the alternative $\Lambda$CDM model
proposed recently by Gubser and Peebles in \cite{Gubser:2004uh}.  We also
comment on the possibility of combining string gases with a period of
cosmological inflation.

In Section VII we conclude.  In the appendices we provide a short review on
conformal transformations and dimensional reduction, necessary for going
between the $10D$ String frame and $4D$ Einstein frames.

In this review, we attempt to provide a comprehensive survey of the existing
SGC literature, focusing on the string theory origin and the importance of
moduli stabilization.  For complementary reviews with emphasis on cosmological
aspects, we refer the reader to
\cite{Brandenberger:2005nz,Brandenberger:2005fb}.

\section{Dynamics of Strings in Time-Dependent Backgrounds \label{section2}}

A closed string in a background generated by its bosonic, massless modes is
described by a nonlinear sigma model \cite{Callan:1985ia}
%
\be \label{sigmamodel} S_{\sigma}=-\frac{1}{4 \pi \alpha^{\prime}}\int
d^{2}\sigma \Bigl( \sqrt{-\gamma} \gamma^{ab}G_{\mu \nu}(X) \;
\partial_{a}X^{\mu}\partial_{b}X^{\nu}+
\epsilon^{ab} B_{\mu \nu}(X)\;
\partial_{a}X^{\mu}\partial_{b}X^{\nu} \Bigr), \ee
%
where $\gamma^{ab}$ is the world-sheet metric, $(2 \pi \alpha^{\prime})$ is the
inverse string tension, $G_{\mu \nu}$ is the background space-time metric,
$B_{\mu \nu}$ is the background antisymmetric tensor. Our convention in this
review will be that coordinates of the full space-time are denoted by $X^\mu$
with $\mu=0 \ldots D-1$, where $D$ is the space-time dimension. Spatial
dimensions parameterized by $X^i$ are denoted by indices $i,j=1 \ldots D-1$,
compact dimensions are given by coordinates $Y^m$ with $m,n$ running over
compact spatial coordinates, and $\sigma^a$ with $\sigma^0 \equiv \tau,
\sigma^1 \equiv \sigma$ are the worldsheet coordinates.

In addition to the action above, one can add a topological term \be
\label{dilaton} S_\phi=-\frac{1}{4 \pi}\int d^{2}\sigma \sqrt{\gamma} \phi(X)
R^{(2)}, \ee where $\phi$ is the background dilaton, which is coupled to the
world-sheet Ricci scalar $R^{(2)}$.  The string coupling is then given in terms
of the vacuum expectation value of the dilaton $g_{s}=e^{ \phi_0}$.

Varying the action (\ref{sigmamodel}) with respect to the fields, $X^{\mu}$,
gives the string equations of motion in a general space-time \bea \label{seom}
\partial_a \Bigl( \sqrt{\gamma} \gamma^{ab} \partial_b X^{\mu}
\Bigr)+\Gamma^{\mu}_{\; \lambda \nu} \sqrt{\gamma}\gamma^{ab}
\partial_a X^{\lambda}
\partial_b X^{\nu} +\frac{1}{2}H^\mu_{\; \lambda \nu}\epsilon^{ab}\partial_a X^{\lambda}
\partial_b X^{\nu} =0.
\eea In addition, one must satisfy the constraint equations \be
\label{constraints} G_{\mu \nu}(X) \Bigl(\partial_a
X^\mu(\sigma,\tau)\partial_b
X^\nu(\sigma,\tau)-\frac{1}{2}\gamma_{ab}\gamma^{cd}
\partial_c X^{\mu} \partial_d X^{\nu}\Bigr)=0.
\ee

The background fields, $G_{\mu \nu}, B_{\mu \nu}$, and $\phi$, are realized as
couplings of the non-linear sigma model as can be seen from the action above.
This model possesses a conformal symmetry classically, but this is spoiled at
the quantum level by anomalies; the couplings evolve in accordance with the
corresponding beta functions\footnote{Actually (\ref{dilaton}) already breaks
conformal symmetry at the classical level, but is none-the-less required
\cite{Fradkin:1984pq}.}. This is equivalent to demanding that the trace of the
world-sheet stress tensor given by
\be
T^{a}_{a}= \beta^{G}_{\mu
\nu}\sqrt{\gamma}\gamma^{ab}\partial_{a}X^{\mu}\partial_{b}X^{\nu}
+\beta^{B}_{\mu
\nu}\epsilon^{ab}\partial_{a}X^{\mu}\partial_{b}X^{\nu}+\beta^{\phi}\sqrt{\gamma}R^{(2)},
\ee
vanishes, where the $\beta$ functions are found, (e.g. by the background field
method) to be \cite{Callan:1985ia}
\bea \label{betaeqns} & & \beta^{G}_{\mu \nu}= \Bigl( R_{\mu \nu} +2
\nabla_{\mu} \nabla_{\nu}\phi -\frac{1}{4} H_{\mu \kappa \sigma} H_{\nu}^{\;
\kappa \sigma}\Bigr) + {\cal O}(\alpha^{\prime}),\nonumber \\ & &
\beta^{B}_{\mu \nu}=\Bigl( \nabla^{\kappa}H_{\kappa \mu \nu} -
2\nabla^{\kappa}\phi H_{\kappa \mu \nu }\Bigr)+ {\cal
O}(\alpha^{\prime}),\nonumber \\ & & \beta^{\phi}= \frac{1}{\sdrs}\Bigl(
\frac{D-26}{48 \pi^2} \Bigr)+\Bigl( 4\nabla_{\kappa}\phi \nabla^{\kappa}
\phi-4\nabla_{\kappa}\nabla^{\kappa}\phi -R +\frac{1}{12}H_{\kappa \mu \nu}
H^{\kappa \mu \nu}\Bigr) + {\cal O}(\alpha^{\prime}), \eea
with $H=dB$ denoting the field strength associated with the field $B_{\mu
\nu}$. Keeping terms {\bf tree level} in $\alpha^{\prime}$, these equations of
motion can be derived from those of the low energy effective action of
supergravity in $D$ space-time dimensions
\be
\label{theaction} S_0=\frac{1}{2 \kappa_D^2}\int d^{D}x \sqrt{-G} e^{-2
\phi}\Bigl(R+c+4(\nabla \phi)^{2}-\frac{1}{12}H^{2} \Bigr) \, , \ee
where $c$ vanishes in the critical case, $D=26$ $(D=10)$ for the bosonic
(super) string, and acts as a cosmological constant in the noncritical case. In
the case $D\leq10$, the prefactor takes the form $2\kappa_D^2=(2 \pi
\sdr)^{D-2}g_s^2(2\pi)^{-1}=16 \pi G_{D}$ with $l_s=\sdr$ the string length and
$G_D$ the $D$ dimensional Newton constant. By noting this prefactor we see that
this action is not only tree level in $\sdrs$, but it is also tree level in
$g_s=e^{\phi_0}$ where $\phi_0$ is the expectation value of the dilaton
\footnote{Higher $g_s$ corrections would come from considering corrections to
the $\beta$ equations from higher genus surfaces for the string world-sheet
corresponding to string interactions (we implicitly used a sphere, genus
zero).}.

The above action exhibits a new symmetry, scale factor duality, that is not
found in pure general relativity.  To see this, let us consider cosmological
solutions, ignoring flux for the moment and working in the critical dimension
($c=0$).We take the metric and dilaton to have the form \bea \label{themetric}
ds^2=-dt^2+\sum_{i=1}^d a_i^2(t)\,dx_i^2, \nonumber \\ a_i \equiv
e^{\lambda_i(t)}, \;\;\;\; \phi=\phi(t), \;\;\;\; d=D-1, \eea where the spatial
directions are taken to be toroidal. It will prove useful to perform a field
redefinition and introduce the shifted dilaton, \be \label{shiffy}
\varphi=2\phi-\ln V=2\phi-\sum_{i=1}^d \lambda_i. \ee Plugging this ansatz for
the fields into the action (\ref{theaction}) one finds that the action is
invariant under the transformation
\be
a_i \rightarrow \frac{1}{a_i}, \;\;\;\;\;\; \lambda_i \rightarrow -\lambda_i,
\;\;\;\;\;\;  \varphi \rightarrow \varphi. \ee This symmetry is known as scale
factor duality, and has interesting consequences for cosmology.  In particular,
it tells us that the effective field theory of dilaton gravity for a small
scale factor is equivalent to that for a large scale factor.

In addition to the low-energy action for the massless modes above, one may
consider the addition of classical or quantum {\em string} matter. One approach
that was first advocated in
\cite{Dabholkar:1995nc,Dabholkar:1990yf,Dabholkar:1989jt} is to include the
action (\ref{sigmamodel}) as a phenomenological matter source for the
background fields in (\ref{theaction}). There are many interpretations of what
such a term may represent. In the supergravity (SUGRA) solutions presented in
\cite{Dabholkar:1995nc,Dabholkar:1990yf,Dabholkar:1989jt}, the authors observed
that the string source was required at the origin to complete the solution.  It
has also been argued that this action can be added as a method for taking into
consideration quantum corrections coming from higher genus worldsheets (see
e.g. \cite{Tseytlin:1991ss,deAlwis:1996ze}). This interpretation is clear from
the additional power of $g_s^2$ that appears in front of the action
(\ref{sigmamodel}) relative to (\ref{theaction})\footnote{The action
(\ref{theaction}) carries a multiplicative factor of $g_s^{-2}$, whereas the
action (\ref{sigmamodel}) has prefactor $g_s^0$.  Thus, the latter is one
higher order in the closed string coupling $g_s^2$ and is related to the 1-loop
free energy coming from strings on a toroidal worldsheet (see for example
\cite{Bassett:2003ck,Borunda:2003xb}).}.

If we consider a single string source for the background fields the total
action becomes, \be \label{totalaction} S=S_0+S_\sigma. \ee Varying this action
we recover the equation of motion of the string (\ref{seom}), the constraints
(\ref{constraints}), and the background equations sourced by the string which
take the form \bea \nonumber & & R_{\mu \nu} +2  \nabla_{\mu} \nabla_{\nu}\phi
-\frac{1}{4} H_{\mu \kappa \sigma} H_{\nu}^{\; \kappa \sigma}=-\frac{\kappa^2_D
e^{2 \phi}}{2 \pi \alpha^{\prime} \sqrt{-G}} \int d^2\sigma \sqrt{\gamma}
\gamma^{ab}
\partial_a X^\mu \partial_b X^\nu \delta^{(D)}(x-X(\sigma)), \\ \label{heqns1}\\ \label{heqn0} & &
\nabla_{\mu}\left( e^{-2\phi} H^{\mu \nu \rho} \right)= \frac{\kappa_D^2}{\pi
\sdrs \sqrt{-G}}\int d^2\sigma \epsilon^{ab}\partial_a X^\nu \partial_b X^\rho
\delta^{(D)}(x-X(\sigma)),\\ & & 4\nabla_{\kappa}\phi \nabla^{\kappa}
\phi-4\nabla_{\kappa}\nabla^{\kappa}\phi -R +\frac{1}{12}H_{\kappa \mu \nu}
H^{\kappa \mu \nu}=0. \label{heqns} \eea From (\ref{heqns1}) we see that the
stress-energy tensor of a single string is \be \label{stress} T_{\mu
\nu}\equiv\frac{2}{\sqrt{-G}}\frac{\delta S_\sigma}{\delta G_{\mu \nu}}
=-\frac{1}{2 \pi \sdrs \sqrt{-G}}\int d^2\sigma \sqrt{\gamma} \gamma^{ab}
\partial_a X^\mu \partial_b X^\nu \delta^{(D)}(x-X(\sigma)). \ee These
equations, along with (\ref{seom}) and (\ref{constraints}), represent a system
of a single string in the presence of its background massless modes. Solving
these equations would seem extremely difficult given the non-linearity of the
problem. However, static solutions were found some time ago
\cite{Dabholkar:1995nc,Dabholkar:1990yf,Dabholkar:1989jt} and these so-called
F-string solutions were shown to preserve some supersymmetries and exhibit
BPS-like properties similar to solitons.  In particular, two parallel strings
satisfy a no-force condition, since the gravitational attraction is canceled by
the scalar exchange of the dilaton and flux. Instead, in SGC we will be
interested in solutions generated by a gas of strings at finite temperature and
in cosmological (time-dependent) background fields.

\section{Cosmology with String Gases \label{sec3}}
We now want to attempt to solve for the background fields allowing for
conditions indicative of early universe cosmology.  As mentioned in the
previous section, generically the equations resulting from (\ref{seom}) are
very difficult to solve.  However, by invoking some approximations that are not
in conflict with cosmological observation, the equations can be made tractable.
We will now explicitly state these approximations leaving a discussion of their
limitations to follow.

\subsection{Assumptions of the string gas approach}
\bi
\item {\bf Homogeneous Fields}: $\;$ We will assume that the background
fields (i.e. metric, flux, and dilaton) are homogeneous and therefore at most
functions of time.  The generalization to inhomogeneous fields was addressed in
\cite{Watson:2003uw,Watson:2004si,Battefeld:2005wv}.

\item  {\bf Adiabatic approximation}: $\;$  We will assume that the background fields
are evolving slowly enough that higher derivative corrections, i.e. $(\sdrs)$
corrections, can be ignored. This means that locally, string sources won't be
influenced by the expansion and their evolution can be characterized by their
scaling behavior.

\item {\bf Weak Coupling}: $\;$  We will work in the region of weak
coupling (i.e. $g_s \ll 1$), and we will choose initial conditions for the
dilaton that preserve this condition. Thus, higher orders corrections in $g_s$,
can be neglected.

\item {\bf Toroidal Spatial Dimensions}: $\;$  We assume that all
spatial dimensions are toroidal and therefore admit non-trivial one cycles.
In the past this assumption was believed to be crucial, however it was later
shown that this condition may be relaxed in some cases, allowing for more phenomenologically
motivated backgrounds \cite{Easther:2002mi}.
\ei

From the point of view of cosmology, all of these approximations are familiar.
However, both the adiabatic and weak coupling approximation are very
restrictive from the string theory perspective.  The string corrections that we
are choosing to ignore may be very important for early universe cosmology,
especially near cosmological singularities. The motivation here is to take a
modest approach by slowly turning on {\em stringy} effects, as one extrapolates
the known cosmological equations backward in time to better understand the
departures from standard big-bang cosmology.  This is to be contrasted to
models of string cosmology that invoke supersymmetry to avoid higher order
corrections.  From the cosmological standpoint, one could argue that these
models are less realistic since supersymmetry should not be expected to hold in
conditions favorable to the early universe, i.e. time-dependent, finite
temperature backgrounds.  It is certainly premature to claim one has a well
established understanding of string theory in cosmological backgrounds, but one
hopes by the SGC approach to better understand what role strings play in the
early universe.

\subsection{Energy and pressure of a string gas}
Given the assumptions stated above, we now want to find cosmological solutions
to the equations (\ref{heqns1})-(\ref{heqns}), given the presence of a string
gas. The time-dependent background fields are \bea ds^2=-dt^2+\sum_{i=1}^d
a_i^2(t)\,dx_i^2, \nonumber \\ a_i \equiv e^{\lambda_i(t)}, \;\;\;\;
\phi=\phi(t), \;\;\;\; B_{\mu \nu}=0, \;\;\;\; d=D-1. \eea

The adiabatic approximation implies that the local effects of expansion on the
string can be neglected, allowing us to simplify the string equation of motion
(\ref{seom}) to \bea
\partial_a \Bigl( \sqrt{\gamma} \gamma^{ab} \partial_b X^{\mu}
\Bigr)+\Gamma^{\mu}_{\; \lambda \nu} \sqrt{\gamma}\gamma^{ab}
\partial_a X^{\lambda}
\partial_b X^{\nu}  \approx \left( \partial_\tau^2- \partial_\sigma^2
\right) X^{\mu}(\sigma,\tau)=0, \eea where we have fixed the gauge of the
worldsheet metric to $\gamma_{ab}=f(\tau, \sigma)\eta_{ab}$, with
$\eta_{ab}=diag (-1,1)$. In this gauge the constraints (\ref{constraints})
become \bea \label{c1} && G_{\mu \nu}\left( \dot{X}^\mu
\dot{X}^\nu+\acute{X}^\mu \acute{X}^\nu \right)=0 \\ \label{c2} && G_{\mu
\nu}\dot{X}^\mu \acute{X}^\nu=0, \eea where $\dot{X}\equiv
\partial_\tau X$ and $\acute{X}\equiv
\partial_\sigma X$. Since the $X^\mu$ satisfy a free wave
equation, their solution can be decomposed into left and right movers \bea
X^\mu=X^\mu_L(\tau+\sigma)+X^\mu_R(\tau-\sigma), \nonumber \\
X^\mu_R=x_R^\mu+\sqrt{\frac{\sdrs}{2}}p_R^\mu(\tau-\sigma)+i\sqrt{\frac{\sdrs}{2}}\sum_{n\neq
0}\frac{1}{n} \alpha_n^\mu e^{-in(\tau-\sigma)}, \nonumber \\ X^\mu_L=
x_L^\mu+\sqrt{\frac{\sdrs}{2}}p_L^\mu(\tau+\sigma)+i\sqrt{\frac{\sdrs}{2}}\sum_{n\neq
0}\frac{1}{n} \tilde{\alpha}^\mu_n e^{-in(\sigma+\tau)}, \eea where $x_R$ and
$x_L$ are the center of mass position, $p_R^\mu$ and $p_L^\mu$ are the center
of mass momentum, and $\alpha$ ($\tilde{\alpha}$) after quantization are the
operators associated with right (left) moving oscillations of the string. If we
take some of the spatial dimensions to be compact with coordinates $Y^m$, then
the center of mass momenta become \bea \label{lr} p_R^m=
\frac{\sdr}{R}\;n^m-\frac{R}{\sdr}\;\omega^m,\nonumber \\ p_L^m=
\frac{\sdr}{R}\; n^m+\frac{R}{\sdr}\;\omega^m, \eea where $R$ is the scale
factor in the $m$th compact direction, $n_m$ is an integer giving the charge of
the Kaluza-Klein momentum in that direction and $\omega^l=G^{mn}\omega_m$ is an
integer giving the winding number of the wound string (Note: The placing of the
indices is important, for a generic metric $G_{mn}$, $n^m$ and $\omega_m$ are
{\em not} integers.). It is important to note that we are again invoking the
adiabatic approximation, since we are treating the scale factor $R$ as a
constant (locally).

If we now substitute this solution into the constraint equation (\ref{c1}) and
use the gauge choice $X^0=E\tau \sdr$ we find\footnote{There is a subtlety here
involving the quantization procedure and obtaining the physical degrees of
freedom.  The correct way to deal with the constraints is to introduce
light-cone coordinates in target space and this results in only the oscillators
in the transverse directions being excited.  We will take this for granted in
what follows and we refer the reader to \cite{Green:1987sp,Green:1987mn} for
details.} \bea \label{genergy} -G_{00}\dot{X}^0 \dot{X}^0\equiv \sdrs
E^2&=&G_{ij} \left(\dot{X}^i \dot{X}^j+\acute{X}^i \acute{X}^j \right) +G_{mn}
\left(\dot{Y}^m \dot{Y}^n+\acute{Y}^m \acute{Y}^n \right)\nonumber \\ &=&\sdrs
\vec{P}^{\; 2}+\frac{(p_L^2+p_R^2)}{2}+2\left( N_L+N_R+a_L+a_R\right), \eea
which is the mass shell condition for the string $E^2=\vec{P}^{\; 2}+M^2$. The
constants $a_R$ and $a_L$ have been added to account for normal ordering, with
$a_R=a_L=-1$ for the bosonic string.  For the heterotic string $a_L=-1$ and $a_R=-\frac{1}{2}$ for the
Neveu-Schwarz sector, while $a_R=0$ for the Ramond-Ramond sector, 
and $N_R$ ($N_L$) is the excitation number of the right
(left) oscillators. We have also allowed for the presence of non-compact
dimensions $X^i$, for which the string has center of mass momentum $\vec{P}$
and we have used $X^i=\sdr \vec{P} \tau$. Using the other constraint (\ref{c2})
we find the level matching condition \be \label{lmatch} p_L^2-p_R^2=4 n_m
\omega^m=4(N_R-N_L+a_R-a_L). \ee From the string mass spectrum we immediately
see that strings are invariant under the same duality as the massless
background fields.  Namely, the string spectrum is unchanged under the
transformation
\be
\frac{R}{\sdr} \rightarrow \frac{\sdr}{R} \;\;\;\; n \longleftrightarrow
\omega, \ee which suggests that strings at small scale factor $\frac{1}{R}$,
behave the same as strings at large scale factor $R$.  This property of the
spectrum, known as {\em t-duality}, is a very important property of strings and
suggests that their effects on cosmological backgrounds may differ greatly from
that of ordinary point particles \cite{Brandenberger:1988aj}.

We would now like to reconsider the stress energy tensor (\ref{stress}) for
this string configuration (see e.g. \cite{deVega:1995bq}). The $T^{00}$
component is given by \bea T^{00}&=&-\frac{1}{2 \pi \sdrs \sqrt{-G}}\int
d^2\sigma \sqrt{\gamma} \gamma^{ab} \partial_a X^0 \partial_b X^0
\delta^{(D)}(x^\mu-X(\sigma)^\mu)\nonumber \\ &=&\frac{1}{2 \pi \sdrs
\sqrt{-G}}\int d^2\sigma \left( \dot{X}^0 \dot{X}^0-\acute{X}^0 \acute{X}^0
\right) \delta^{(D)}(x^\mu-X^\mu(\sigma)), \eea where we have again used the
conformal gauge for the worldsheet metric $g_{ab}=f(\tau,\sigma)\eta_{ab}$.
Noting our previous choice of $X^{0}=\sdr E \tau$, we find \bea
T^{00}&=&\frac{1}{2 \pi \sdrs |\dot{X}^0| \sqrt{-G} }\int^{2\pi \sdr}_0 d\sigma
\; \delta^{(D-1)}(x^i-X^i(\sigma)) \left( \dot{X}^0 \dot{X}^0-\acute{X}^0
\acute{X}^0 \right)_{\tau=\tau(X^0)}\nonumber \\ &=&\frac{E}{\sqrt{-G_{{D-1}}}}
\;\delta^{(D-1)}(x^i-X^i(\sigma)), \eea which is the energy density of a single
string with the delta function enforcing that there is no contribution unless
we are at the position of the string. The explicit formula for the energy of
the string in terms of its oscillations and momentum then follows from
(\ref{lr}) and the constraint (\ref{genergy}) \be \label{tstringenergy}
E=\sqrt{\vec{P}^{\; 2}+G^{mn}\left( n_m+\frac{\omega_m}{\sdrs}  \right) \left(
n_n+\frac{\omega_n}{\sdrs}  \right)+\frac{4}{\sdrs}\left( N_L+a_L\right)}, \ee
where we have eliminated $N_R$ in favor of the other quantum numbers by using
(\ref{lmatch}). It is straight forward to generalize this to a gas of $N$
strings. We simply average over the delta function sources and the energy
density of the string gas is \be \label{rho} \rho=\sum_s \tilde{n}_s E_s, \ee
where the sum is over all species $s$ and $\tilde{n}_s={N_s}{V}^{-1}$ is the
number density of the string gas in spatial volume $V$, with a particular set
of quantum numbers $n, \omega, N_L, N_R$.  We will assume that the gas is a
perfect fluid and non-interacting. Therefore, we can find the pressure in the
$i$-th direction \be \label{p} p_i=-\frac{1}{V}\frac{\partial (\rho V)}{
\lambda_i}, \ee where $a_i=\exp(\lambda_i)$ is the scale factor in the $i$-th
direction.

As a simple example, let us consider two bosonic string gases ($a_L=a_R$)
composed of strings wrapping the compact dimensions ($\omega_m \neq 0,
n_m=N_L=N_R=0$) and strings with momentum in the compact dimensions ($n_m \neq
0, \omega_m=N_L=N_R=0$). Assuming we can neglect the non-compact momenta,
$\vec{P}=0$, their energy density and pressure are given by \bea \label{epw}
\rho_w&=&\sum_{l=1}^d \tilde{n}_w^{(l)} e^{\lambda_l(t)}, \;\;\;\;\;\; \rho_m=
\sum_{l=1}^d \tilde{n}_m^{(l)} e^{-\lambda_l(t)},\nonumber \\
p_w^{(l)}&=&-\tilde{n}_w e^{\lambda_l(t)}, \;\;\;\;\;\;\;\;\;\;
p^{(l)}_m=\tilde{n}_m e^{-\lambda_l(t)}, \eea where we now take $d$ to denote
the number of compact directions and we have vanishing pressure in the $D-1-d$
non-compact dimensions. We have lifted the scale factors $R_i=e^{\lambda_i}$ to
time-dependent functions using the adiabatic approximation, with $V(t)$ the
time-dependent spatial volume ($\ln V(t)=\sum_{i=1}^{D-1} \lambda_i(t)$). For
simplicity we have absorbed the winding and momentum numbers $\omega, n$ into
the number density of winding and momentum modes in the $l$th direction
$\tilde{n}_w^{(l)}$ and $\tilde{n}_m^{(l)}$.  We have also renormalized the
mass to remove the tachyonic zero point energy $a_L$, which would be
automatically removed in the case of heterotic strings. Here we do this by
hand, since we are mainly concerned with the scaling of the string energy with
$\lambda_i$.  Given an isotropic distribution of strings in the extra
dimensions, we find that the equation of state for winding and momentum modes
is \be \label{eosm} p_w=-\frac{1}{d}\rho_w, \;\;\;\;\;\;\;\;\;\;
p_m=\frac{1}{d}\rho_m, \ee respectively.  We see that winding modes contribute
negative pressure whereas the momentum modes scale as radiation filling the
extra dimensions. To close this section, we have found that under the
assumption that the string gas can be modeled as a perfect fluid, the stress
energy tensor of a single string (\ref{stress}) can be generalized to \be
\label{gasstress} T^{\mu}_{\nu}=diag\left(-\rho, p_1, p_2, \ldots , p_{D-1}
\right), \ee where the energy density and pressure are given by (\ref{rho}) and
(\ref{p}), respectively.

\section{Classical dynamics of string gases}

\subsection{Initial Conditions and the Dimensionality of
Space-time} One of the successes of SGC is the possibility to explain the
emergence of three large and isotropic spatial dimensions, while six remain
stabilized near the string scale.  In this way, SGC is the only cosmological
model thus far that has attempted to explain the dimensionality of space-time
dynamically\footnote{ However, for recent variations of the ideas to be
discussed see \cite{Majumdar:2002hy,Karch:2005yz,Durrer:2005nz}.}. The
qualitative argument, due to Brandenberger and Vafa
\cite{Brandenberger:1988aj}, was that winding string modes can maintain
equilibrium in at most three spatial dimensions.  This is based on the simple
fact that $p$ dimensional objects can generically intersect in at most $2p+1$
dimensions and the intuition that string interactions are due to intersections.
They argued that once the winding modes annihilate with anti-winding modes,
three spatial dimensions would be free to expand while the remaining six should
remain confined by winding modes near the string scale. Winding modes were
shown to possess such confining behavior quantitatively in
\cite{Tseytlin:1991xk}. There, the importance of the dilaton was stressed
because this led to the observation that the negative pressure of winding modes
leads to contraction rather than accelerated expansion.  It was later observed
that the dilaton is not the critical feature restoring the Newtonian intuition
per se, rather it is the effect of anisotropies\footnote{An easy way to see
this is to think of the dilaton as the scale factor of another $11$th
dimension. Thus, instead of the dilaton, one could simply take one of the other
scale factors to evolve anisotropically while keeping the dilaton; this would
still lead to the same conclusions as in \cite{Tseytlin:1991xk}.}
\footnote{We thank Amanda Weltman and Brian Greene for discussions on this point.}. 
In fact, it was shown many years ago that string winding modes
could lead to confinement in the case of general relativity
\cite{Kripfganz:1987rh}.

This counting argument was verified numerically in a static background,
focusing on cosmic strings in \cite{Sakellariadou:1995vk} (see also
\cite{Cleaver:1994bw} ) and later extended to the case of branes in
\cite{Alexander:2000xv}, where it was argued that the strings remain the
important objects, since branes fall out of equilibrium sooner than strings,
leading to a hierarchial structure of dimensions. The setup has been
generalized to more complex topologies
\cite{Easther:2002mi,Easson:2001fy,Biswas:2003cx} and many authors elaborated
on these basic arguments
\cite{Kaya:2003py,Rador:2005ib,Rador:2005we,Rador:2005vq,Kaya:2003vj,Kaya:2004yj,Arapoglu:2004yf,Kaya:2005qb,Kaya:2005qm,Kim:2004ca,Park:1999xn,Hotta:1997yh,Deo:1991mp,Deo:1991af}.
The t-duality of branes was discussed in the context of SGC in
\cite{Boehm:2002bm}.  Other recent attempts to address dimensionality making
use of branes in a different way have appeared in
\cite{Durrer:2005nz,Karch:2005yz}.

Despite the appeal of the BV argument, there remain serious challenges for its
quantitative realization.  As a first step, a study in eleven dimensional
supergravity \cite{Easther:2002qk} employing a fixed wrapping matrix (based on
the counting argument) yielded indeed the predicted anisotropic expansion.
However, the internal dimensions were not stabilized, but simply grew slower.
This work was extended in \cite{Easther:2003dd} by studying the coupled
Einstein-Boltzmann equations for a thermal brane gas. It was found that only
highly fine tuned initial conditions yield the desired outcome. The most recent
study \cite{Easther:2004sd} focusing on dilaton gravity confirmed these
results: either all dimensions grow large, since the string gas annihilated
entirely, or all dimension stay small, since the string gas froze out --
intermediate solutions can only be achieved by fine-tuning the initial
conditions. It was also observed that a string gas freezes out quite quickly
due to the coupling to the rolling dilaton \cite{Danos:2004jz}.

A crucial input is the interaction rate of strings \cite{Polchinski:1988cn}
that lead to the corresponding Boltzmann equation.
The interaction probability relies on the value of $g_s$ and therefore the
dilaton.  As the dilaton runs to weak coupling this means that interaction
probabilities go to zero.  Secondly, viewing interactions as intersections is
an entirely classical argument\footnote{We thank Liam McAllister for
discussions.}. At the level of supergravity one has exchange of closed strings
that mediate interactions.  This increases the probability of interaction,
since closed string exchange can take place in any number of dimensions with
the only {\em dilution} being due to the force following a generalized Newton
law, i.e. $F \sim \frac{1}{r^{D-2}}$. Henceforth, the conclusion of Easther
et.al.'s investigations have been that compactification of all or none of the
dimensions is the most probable configuration
\cite{Easther:2004sd,Easther:2002qk,Easther:2002mi}. However, this analysis was
done given our rather limited knowledge of string theory dynamics.  In
particular, our knowledge of cosmological solutions when all radii are taken to
be at the string scale is sketchy at best.  A more complete knowledge of both
curvature corrections ($\sdrs$) and the strong coupling behavior of the theory
could certainly change this outcome.  Moreover, time dependent solutions of the
full string theory continue to be an avenue that is being actively pursued. It
will be interesting to see if the BV argument will hold, given a more complete
understanding of string theory dynamics.  While awaiting this progress, we will
simply assume in what follows that winding modes were able to annihilate in
three spatial dimensions, causing those to be free to expand while a winding
mode gas remains in the other six. Thus, our initial conditions will be no more
unnatural than those of usual models of cosmology.

\subsection{Cosmological Evolution in the Presence of a String Gas}
Anticipating the $D=3+1+6$ split, due to the annihilation of the winding modes
in three dimensions, let us consider the following background field
configuration \bea ds^2=-dt^2+e^{2 \lambda(t)} d\vec{x}^2+e^{2 \nu(t)}
d\vec{y}^2,\\ H_3=h \; dx^1 \wedge dx^2 \wedge dx^3, \;\;\;\;\;\; \phi=\phi(t),
\eea where $H_3$ is a constant three form flux restricted by the expected
symmetries (namely, $3+1+6$)\footnote{More general flux configurations were
considered in \cite{Brandenberger:2005bd,Campos:2005da,Kanno:2005ck} where it
is clear that there is still much to consider.}. We want to consider these
background fields in the equations of motion (\ref{heqns1})-(\ref{heqns}), with
the string sources replaced by the averaged stress tensor of the string gas
(\ref{gasstress}). We have \bea \label{e1} & & R_{\mu \nu} +2 \nabla_{\mu}
\nabla_{\nu}\phi -\frac{1}{4} H_{\mu \kappa \sigma} H_{\nu}^{\; \kappa
\sigma}=16 \pi G_{10} e^{2 \phi} T_{\mu \nu}, \\ \label{e2} & &
\nabla_{\mu}\left( e^{-2\phi} H^{\mu \nu \rho} \right)=0,\\ &
&4\nabla_{\kappa}\phi \nabla^{\kappa} \phi-2\nabla_{\kappa}\nabla^{\kappa}\phi
-\frac{1}{6}H_{\kappa \mu \nu} H^{\kappa \mu \nu}=16 \pi G_{10} e^{2 \phi} T,
\label{e3} \eea where $G_{10}$ is the ten-dimensional Newton constant, $T
\equiv T^{\mu}_\mu$ is the trace of the stress tensor, and we have used the
trace of (\ref{e1}) to rewrite the last equation. Writing (\ref{e3}) in this
form allows us to make the important observation that (ignoring flux) the
dilaton can only evolve if matter is not conformal (i.e. $T \neq 0$). This
condition will be respected by string gases in general, and it is this
important observation that makes string cosmology (dilaton gravity) very
different from ordinary cosmology. The flux equation (\ref{e2}) is trivially
satisfied given the ansatz for the background fields and we assume that the
flux of the strings themselves average to zero\footnote{The vanishing of the
total flux is required for consistency on the compact space, however local
sources can prove interesting in a time-dependent background
\cite{Brandenberger:2005bd}.}. The remaining equations can be written in the
form \bea \label{eom} c-3 \ld^2-6\nd^2+\fd^2-\frac{h^2}{2}e^{-6
\lambda}=e^\varphi E , \\ \ldd-\fd
\ld-\frac{1}{2}h^2e^{-6\lambda}=\frac{1}{2}e^\varphi P_3 \label{eom2} \\
\label{eom3} \ndd-\fd \nd=\frac{1}{2} e^\varphi, P_6,\\ \label{eoml} \fdd-3
\ld^2-6 \nd^2=\frac{1}{2}e^{\varphi} E, \eea where we have introduced the
energy $E=\rho V$, we define the scaled pressure $P_i=p_i V$, and $\varphi$ is
the shifted dilaton (\ref{shiffy}). The first equation is an energy constraint,
which if satisfied at some initial time will remain satisfied for all times.
The sources obey a conservation equation,
\be
\dot{E}+3\ld P_3+6 \nd P_6=0. \ee From the above equations we see that the term
involving flux will be negligible at late-times, since it scales as $a^{-6}$.
However, in the early universe as one approaches the cosmological singularity
this term may become vital to understanding the dynamics \cite{Friess:2004zk}.
Also, if we decide to work in the non-critical theory (i.e. $c \neq 0$), we see
that $c$ acts as an effective cosmological constant.

\subsection{Summary of 10d Dynamics and Moduli Stabilization}
We will now briefly review the results of various authors in studying the
system of equations (\ref{eom})-(\ref{eoml}), where it will be assumed that
$h=0$ and $c=0$ unless noted otherwise. In
\cite{Brandenberger:2001kj,Easson:2001re}, the above equations were studied
with energy and pressure given by a gas of string winding modes as in
(\ref{epw}).  There it was shown that the universe remains in a period of
cosmological loitering until all of the winding modes have annihilated.  Once
the winding modes have all annihilated the dimensions are freed to grow large.
It was observed that the period of loitering would resolve the horizon problem,
without the need to invoke cosmological inflation. These results agree with the
earlier study in \cite{Tseytlin:1991xk}, where it was shown that the negative
pressure of the string winding modes leads to contraction in string cosmology,
{\em not} inflation.  In \cite{Watson:2002nx}, the effect of the winding mode
annihilation processes on the three dimensions growing large was shown to lead
to a natural explanation for the observed isotropy of our universe.  This
resulted from the annihilation rate depending on the size of the dimension and
the expansion rate depending on the number of winding modes present. Moreover,
the string winding modes annihilate into unwound closed string loops, or
momentum modes, which we saw from (\ref{eosm}) scale as radiation (d=3 in this
case). Thus, it was shown that a large, three dimensional, radiation dominated
universe naturally evolves from SGC.

\begin{figure}[!]
\includegraphics[totalheight=3 in,keepaspectratio]{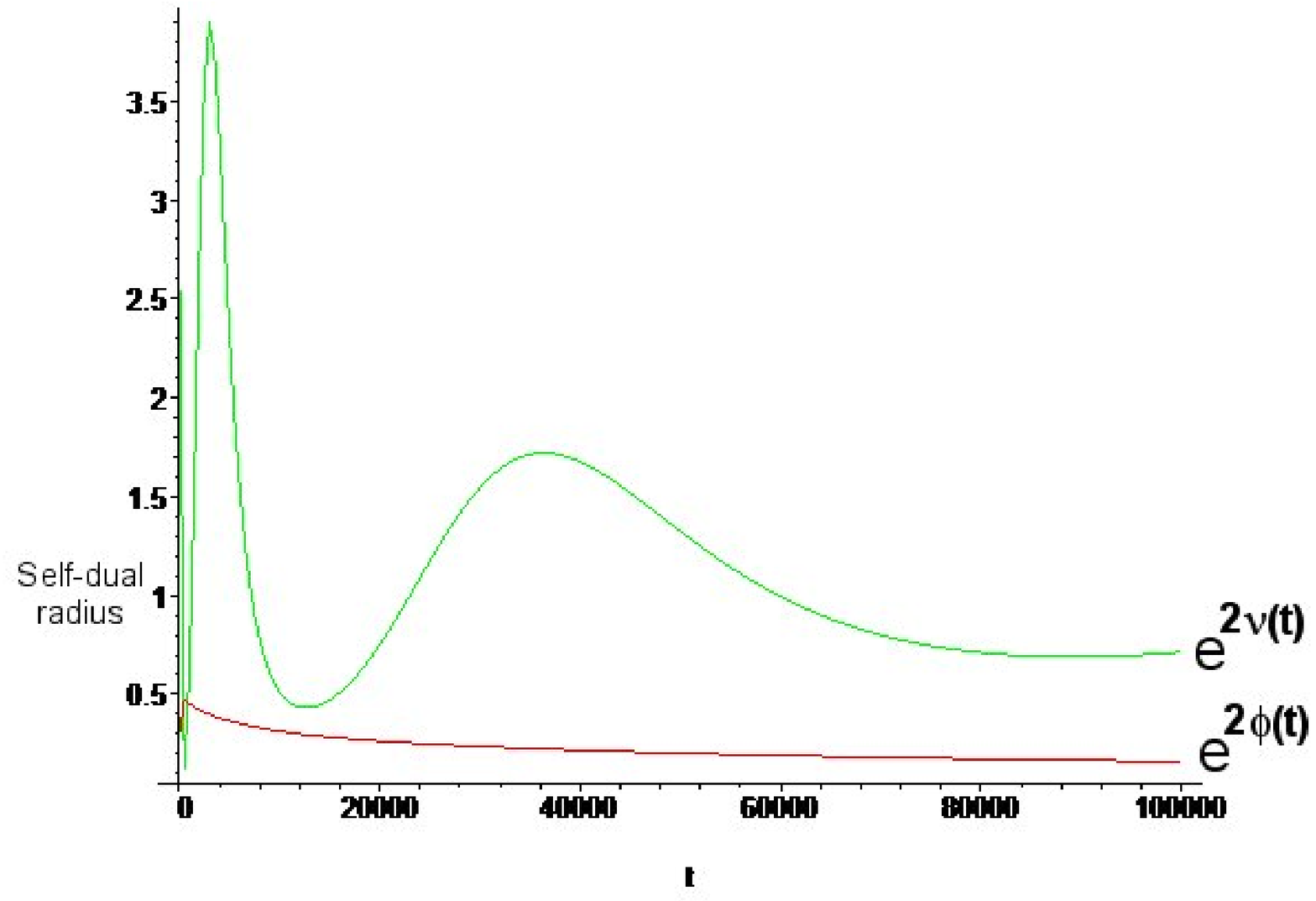}
\caption{This graph shows the primary results of \cite{Watson:2003gf}, where
stabilization of the string frame radion (green or light line) in the presence
of string winding and momentum modes was demonstrated. The damping of the
oscillations relied crucially on the dilaton running slowly to weak coupling
(red or dark line).} \label{fig1}
\end{figure}

In the above investigations, the stabilization of the other six dimensions was
assumed {\em a priori}.  In \cite{Watson:2003gf}, these dimensions were
included and filled with a gas of string winding modes and a gas of string
momentum modes, with energy and pressure as in (\ref{epw}).  It was shown that
as the three spatial dimensions continue to grow large, the six compact
dimensions will oscillate about the self-dual radius, since winding modes were
unable to annihilate in these dimensions via the BV argument discussed above.
The oscillations are the result of the negative pressure of the string winding
modes ($p_w=-\tilde{n}e^{\nu}$) driving the radius to smaller values and the
positive pressure of the string momentum modes ($p_m=\tilde{n}_m e^{-\nu}$)
driving the radius to larger values. For an equal number of winding and
momentum modes (i.e. $\tilde{n}_w=\tilde{n}_m$) one finds that the evolution is
driven to the critical radius, the so-called self-dual radius $\nu=0$ or
$b=\sdr$ where the total pressure vanishes and t-duality is
restored\footnote{Similar results were reported by Tseyltin sometime ago,
however no details regarding the anisotropic case were given in
\cite{Tseytlin:1991ss}.}.  In order for stabilization to occur it was crucial
that the dilaton ran to weak coupling.  This running of the dilaton leads to a
damping effect of the internal dimensions, as can be seen in Figure \ref{fig1}.
The running of the dilaton implies that the Newton constant will evolve and
this will prove problematic at late-times.  However, the important point here
is that during the early stages of the evolution, the extra dimensions are
naturally led to the self-dual radius. In fact, the important point stressed in
\cite{Watson:2003gf} and later elaborated on in
\cite{Patil:2004zp,Watson:2004aq,Patil:2005fi} is the presence of additional
massless string states that become massless at the self-dual radius and should
therefore be considered in the low energy action. We will see in Section
\ref{modulitrap} that these states can have a very important effect resulting
in a stabilization mechanism for the extra dimensions.

So far, we have ignored the problem of inhomogeneities, since we have assumed
all the background fields to be homogeneous. This problem was considered at
late-times in \cite{Watson:2003uw} and \cite{Watson:2004vs}, where it was shown
that the dilaton again plays a vital role. It was found that as long as the
dilaton continues to roll towards weak coupling, perturbations will be under
control and stability of the string frame radion will persist. So it would
appear that the dilaton plays a very important role in SGC, but as we will see
in the next section, it must ultimately be stabilized if SGC is to agree with
observation.
The role of inhomogeneities at early-times is a much more challenging problem.
As we approach the cosmological singularity, we might hope that the finiteness
of strings would resolve the singularity and/or provide a bounce. Although many
different approaches have been attempted (see e.g.
\cite{Khoury:2001bz,Gasperini:2002bn}), no convincing models have been found
\cite{Polchinski:2005qi}. This seems a promising area for SGC to investigate,
given the various duality properties exhibited by the string gas and background
fields \cite{Brandenberger:1988aj}.  The dynamics of SGC as the singularity is
approached has been largely ignored due to the lack of control of string
corrections and the expected breakdown of the assumptions stated in Section
\ref{sec3}.  One attempt at understanding the evolution is the work of
\cite{Friess:2004zk}, where it was found that the background flux would play a
crucial role and could no longer be ignored.  It will be interesting to see how
string winding modes and strings as local sources of flux can effect the
evolution towards the singularity. This presents an important challenge for
SGC.

Before closing this section, we would like to briefly mention some other
consideration of SGC dynamics that have appeared in the literature. The
assumption of toroidal geometry used in (\ref{eom})-(\ref{eoml}) was
generalized to orbifold backgrounds in \cite{{Easther:2002mi}}, where it was
found that the confining behavior of winding modes still persists even in the
absence of non-trivial homotopy. Interactions of the string winding and
momentum mode gases were considered in both \cite{Bastero-Gil:2001nu} and
\cite{Danos:2004jz}, where in the former it was argued that correlations
between the winding and momentum modes lead to modified dispersion relations
that may help explain the small value of the cosmological constant. In addition
to the study of the equations (\ref{eom})-(\ref{eom}), attempts to extend SGC
to M-theory via its connection with $11D$ SUGRA was considered in
\cite{Alexander:2002gj,Easther:2004sd,Easther:2002qk,Campos:2004yn}. Campos has
considered the importance of background flux in SGC
\cite{Campos:2003gj,Campos:2003ip,Campos:2005da}. Whereas in
\cite{Brandenberger:2005bd} the effects of strings as sources of flux was
considered, and in particular their ability to stabilize shape moduli in
addition to the radion. The idea of inflation or cosmic acceleration from SGC
was discussed in \cite{Brandenberger:2003ge,Kaya:2004yj,Parry:2001zg} and
remains a difficult challenge for SGC.  We refer the reader to our references
for addition papers on SGC.

\subsection{$4D$ Dynamics and the Effective Potential\label{section:dilatons}}
Thus far the stability analysis of the extra dimensions has been carried out in
the string frame.  In this frame it has been shown that the radion is
stabilized at the self-dual radius by the competing negative and positive
pressure of the stringy matter, along with damping provided by the dilaton
which continues to run to weak coupling. However, at late-times an evolving
dilaton is problematic for both particle phenomenology and moduli
stabilization. In fact, any evolving gravitational scalar will lead to a
changing gravitational constant $G_N$, which is tightly constrained by fifth
force experiments (see e.g. \cite{Gubser:2004uf}).  Moreover, because the
Einstein frame radion is actually a linear combination of the string frame
dilaton and radion, we will find that the extra dimensions will be unstable as
long as the dilaton evolves. We will briefly discuss possibilities for
dynamically stabilizing the dilaton in the next section, but let us first
review the problem of stability as discussed in \cite{Battefeld:2004xw} (see
also \cite{Berndsen:2004tj,Berndsen:2005qq,Easson:2005ug}).

In order to examine the late time behavior of SGC it is most appropriate to
work in the $4D$ Einstein frame. Since we have focused on homogeneous fields,
the {\em physical quantities} originating from these equations are equivalent
to those of the $10D$ string frame we have considered thus far; this is simply
the consistency of dimensional reduction. The $10D$ Einstein frame metric can
be rewritten in terms of the string frame scale factors and dilaton as
\be
ds^2_E=-dt_E^2+e^{\phi/2} a^2(t)d\vec{x}^2+e^{\phi/2} b^2(t)d\vec{y}^2,\;\;\;
\text{with} \;\;\; dt_E^2=e^{\phi/2}dt_s^2 \ee which immediately allows one to
see the problem.  Even if one fixes $b(t)$, the dilaton evolution still
prevents stabilization of the Einstein frame radion.  We see that in this case
the Einstein frame makes this instability manifest in a simple way. However,
the same conclusion could have been reached in the string frame by more
complicated methods, such as identifying the physical radion and examining the
corresponding two-point function. The important point is that {\bf the two
frames are physically equivalent}, but the instability is manifest in the
Einstein frame. In addition to the problem of the dilaton, we will see that
from the $4D$ Einstein frame additional problems arise regarding the dilution
of our string matter as a source of stabilization.

Beginning from the $10D$ string frame action (\ref{theaction}) one can reduce
to the $4D$ Einstein frame by a conformal transformation followed by field
redefinitions to canonically normalize the scalars. We leave the details to the
appendix, where we find \bea \nonumber S_{4}= \int d^4x \sqrt{-{g}} \Biggl(
\frac{1}{16 \pi G} \Bigl[ R[{g}_{\mu \nu}]-\frac{1}{2} {g}^{\mu \nu}
{\nabla}_\mu \psi {\nabla}_\nu \psi-\frac{1}{2} {g}^{\mu \nu}{\nabla}_{\mu}
{\phi} {\nabla}_{\nu} {\phi}\Bigr] \label{effaction4d}
-e^{4\phi-\sqrt{\frac{d}{2}}\psi}
\mathcal{V}^{(4+d)}_s(\lambda,\varphi,\psi)\Biggr). \eea where again we neglect
flux and work in the critical dimensions $c=0$ and where the $4D$ Newton
constant is given by $16 \pi G={2 \pi \sdrs g_s^2}$. The canonically normalized
scalars, $\phi$ and $\psi$ are then $4D$ fluctuations about the fixed values
for the dilaton and radion, respectively. The $10D$ string frame potential
$\mathcal{V}^{(4+d)}_s$ includes the effects of any wrapped branes or strings,
flux, cosmological constant or any other contribution to the energy density.

As a simple example, consider a cosmological constant arising in the $10D$
string frame, such as appears in the RR sector of massive Type II-A
supergravity.  We see that in the $4D$ Einstein frame this term is no longer
constant
\be
\mathcal{V}^{(4+d)}_s \sim \Lambda \;\;\; \longrightarrow \;\;\;
\mathcal{V}^{(4)}_E= \frac{e^{4\phi-\sqrt{\frac{d}{2}}\psi}}{(2\pi\sdr)^4}\,
\Lambda, \ee and if we assume weak coupling, i.e. $\phi \rightarrow -|\phi|$,
we see that one gets a exponential runaway potential.

We would now like to see if the situation improves in SGC, where it seemed
earlier that wrapped strings could stabilize the extra dimensions. We are
interested in potentials coming from wrapped and moving branes and strings on
the compact space.  Assuming the string frame metric to have the form
\be
ds^2=-dt^2+a^2(t)dx^2+b^2(t)dy^2, \ee we can write the $10D$ string frame
potential as
\be
\mathcal{V}^{(4+d)}_s=\mu \frac{N b^k}{a^{3} b^d} \ee where $\mu=(2\pi
\sdr)^{-4}$ and following the notation in the appendix we have absorbed a
factor of $(2\pi \sdr)^{6}$ coming from the compactification into the
definition of $\mathcal{V}_s^{(4+d)}$. The number of strings (branes) is given
by $N$ and $k \leq |d|$ is the type of strings (branes) (e.g., $k=2$ is a wound
2-brane and $k=-1$ is a string with Kaluza-Klein momentum in one compact
direction). Of course, this expression is just a generalization of our earlier
expression (\ref{epw}), for the energy density of winding and momentum string
gases. The reduction to $4D$ leaves the potential unchanged, but we must
transform the scale factor $a(t)$ when moving to the Einstein frame, i.e.
$\tilde{a}(t)=e^{-\varphi} a(t)$, where $\varphi$ is the canonical $4D$ dilaton
$\varphi=2\phi-d \ln b$ and $\tilde{a}(t)$ is the Einstein frame scale factor.
The potential becomes
\be
\mathcal{V}^{(4+d)}_s=\mu \tilde{n} e^{-\frac{3}{2}\varphi}b^{k-d}= \mu
\tilde{n} e^{-3\phi} b^{k+\frac{d}{2}}, \ee where $\tilde{n}$ is the number
density in the Einstein frame and we have expressed the potential in terms of
the unshifted dilaton. Comparing this potential with the action
(\ref{effaction4d}) we find that the potential in the $4D$ Einstein frame is
\be \label{stringpotential} \mathcal{V}^{(4)}_E=\mu \tilde{n} e^{\phi}
b^{k-\frac{d}{2}}=\mu \tilde{n}e^{-|\phi|}\exp \left[\left(
{\frac{2k-d}{2\sqrt{2d}}}\right)\psi\right], \ee where in the last step we have
expressed the radion in terms of the canonical variable $\psi$ and we have
assumed the dilaton evolves to weak coupling. From this potential we can see
that a confining potential only arises if $k\geq \frac{d}{2}$.  For the case of
a winding string ($k=1$) this is only true for a single extra dimension $d=1$
and even then there is an overall factor of the dilaton diluting this
potential. We conclude that a gas of purely winding strings is not enough to
stabilize the extra dimensions.

Given this negative outcome, we would now like to consider a gas composed of a
less restrictive string configuration.  Let us consider the stress energy
tensor for a gas of heterotic strings given by (\ref{rho}), (\ref{p}), and
(\ref{gasstress}). The energy of the individual string is given by
(\ref{tstringenergy}) and in the case of the heterotic string takes the form
\be \label{hetenergy} E_{HE}=\sqrt{G^{mn}\left( n_m+\frac{\omega_m}{\sdrs}
\right) \left( n_n+\frac{\omega_n}{\sdrs}  \right)+\frac{4}{\sdrs}\left(
N_L-1\right)}, \ee and the level matching condition follows from (\ref{lmatch})
as
\be
n_m\omega^m=N_R-N_L+\frac{1}{2}, \ee where we have used $a_L=-1$ and
$a_R=\frac{1}{2}$ for the heterotic string and we have again assumed that
$\vec{P}=0$.  We are interested in ground state configurations of the string,
which in the case of NS Heterotic strings means setting the right oscillators
to their minimum value, $N_R=\frac{1}{2}$ (see \cite{Polchinski:1998rr} for
details).  We then want to consider non-oscillatory states ($N_L=0$), since we
are interested in the terms that contain explicit dependence on the scale
factor of the extra dimensions. With these assumptions the energy and
constraint become \bea \label{thenergy}
E_{(N_L=0,N_R=\frac{1}{2},\vec{n},\vec{\omega})}&=&\sqrt{G^{mn}\left(
n_m+\frac{\omega_m}{\sdrs}  \right) \left( n_n+\frac{\omega_n}{\sdrs}
\right)-\frac{4}{\sdrs}},\nonumber\\ n_m\omega^m&=&1. \eea Let us consider the
energy at the self-dual radius $b=\sdr$, where we have seen that the higher
dimensional evolution naturally led us.  At the self-dual radius, we can see
from the energy and level matching condition that additional massless states
will occur if the winding and momentum numbers satisfy the conditions \be
\label{msc} n \cdot n+\omega \cdot \omega=2, \;\;\;\;\;\;\;\;\; n \cdot
\omega=1, \ee where we introduce the notation $n \cdot n \equiv\delta^{mn}n_m
n_n$, $\omega \cdot \omega\equiv\delta_{mn}\omega^m \omega^n$, and $n \cdot
\omega \equiv n_m\omega^m$ with $\delta_{mn}$ the Kronecker delta symbol. Given
that these states become massless at the self-dual radius and then grow massive
as the radion leaves, one might hope that this could lead to a stabilizing
potential in the $4D$ Einstein frame. Upon reducing we find \bea
\mathcal{V}_s^{(4+d)}&=&\mu \tilde{n} e^{-3\phi} b^{\frac{d}{2}}E \nonumber \\
&=&\mu \tilde{n} e^{-3\phi} e^{\frac{1}{2}\sqrt{\frac{d}{2}}\psi} E, \eea where
we have rescaled $E$ to put all $\sdr$ dependence in $\mu$ for simplicity. The
$4D$ Einstein frame number density is given by $\tilde{n}$, $\phi$ is the
unshifted dilaton, $\psi$ is the normalized radion, and the energy $E$ is given
by \bea E&=&\sqrt{\frac{n \cdot n}{b^2}+\omega \cdot \omega b^2-2 n \cdot
\omega} =\left| \frac{n}{b}-\omega b\right| \nonumber\\ &=&2\left|  \sinh
\left(\frac{\psi}{\sqrt{2d}}\right)\right|, \eea where we have set $n=\omega$
to satisfy the massless state conditions (\ref{msc}). The $4D$ Einstein frame
potential takes the form \be \label{heef} \mathcal{V}_E^{(4)}=2\mu \tilde{n}
e^{\phi -\frac{1}{2}\sqrt{\frac{d}{2}}\psi}\left| \sinh\left(
\frac{\psi}{\sqrt{2d}} \right) \right|. \ee This potential does admit a local
minimum, but as the dilaton runs to weak coupling the minimum becomes shallow.
This result is sensitive to initial conditions, but can lead to interesting
phenomenology if the dilaton is taken into close consideration.

A more serious objection to the above potential comes from considering its
inclusion in the low energy effective action (LEEA).  That is, for $b \neq
\sdr$ we saw that the string states are massive.  In fact, they are very heavy
since their masses are string scale.  Only near the self-dual radius ($b
\approx \sdr$) do these states become light enough that it makes sense to
include them in the LEEA.  One can attempt to avoid this objection by insisting
that by including the strings as sources we have managed to capture the full
action and not just the LEEA.  However, the problem resurfaces if we recall
that we chose a very specific heterotic string gas in order to obtain the
potential (\ref{heef}).  This is simply the objection that if we include one
massive state of the string, don't we have to include all of them? In fact, for
many other states of the heterotic string we find additional points (even
surfaces) in moduli space where the states become light.  These also act as
attractors for the radion and the point one gets trapped at becomes a function
of initial conditions.  We will see in the next section that there is a
possible resolution to the question of the relevance of such trapping
potentials in the LEEA.

\section{Quantum dynamics of string gases \label{modulitrap}}
In the last section we saw that a heterotic string gas carrying both winding
and momentum can result in a stabilizing potential for the string frame radion.
This potential resulted from the dependence of the string mass on the value of
the radion.  The dynamics then drives the radion to values that minimize the
energy of the string gas, which in the case we considered corresponded to the
self-dual radius $b=\sdr$.  This leads to a trapping mechanism for the radion,
given that the string gas survives the cosmological redshift and the dilution
resulting from the running of the dilaton. This idea of trapping by a massive
gas will be referred to as {\em classical trapping}\footnote{This idea has been
considered in other works; including M-theory matrix models
\cite{Helling:2000kz}, flop transitions on the conifold in both the M-theory
\cite{Mohaupt:2004pr}, Type IIA \cite{Mohaupt:2005pa}, and Type IIB string
theory \cite{Lukas:2004du} and for a gas of massive extremal blackholes
\cite{Kaloper:2004yj}.}. The terminology {\em classical} is used here to
signify that this mechanism results from considering the effects of classical
string gas matter sources on the classical dilaton-gravity equations. As we
mentioned in the last section, one serious objection to this idea is that we
have chosen to include states that are very massive at generic locations of the
moduli space, but we have not included all the other massive string states.

An alternative (but not unrelated) point of view is to consider the quantum
production of these states as we pass near places in the moduli space where
additional string states become light. This is the idea of {\em quantum}
trapping \cite{Kofman:2004yc,Watson:2004aq} and differs from the classical case
in that the states are not included in the action initially.  Instead, these
states are produced as the modulus rolls near a place in moduli space where
additional states become massless.  Then, the modulus continues to evolve, but
because the mass of the produced states depends on the modulus, backreaction of
the produced string gas results in a confining potential which can trap the
modulus. It turns out that such points, which we will call Enhanced Symmetry
Points (ESP), are very common in moduli space \cite{Horne:1994mi}. The
ubiquitousness of such states in string models means that we can expect such
trapping to occur as a natural consequence of the dynamics. However, it also
means that the determination of the string vacuum, and thus our universe, may
not be unique.  

To see how {\em quantum} moduli trapping works, let us consider the simple case
of a bosonic string compactification on $\mathcal{M}^4 \times S^1$. Introducing
complex light-cone coordinates on the world-sheet, the string action
(\ref{sigmamodel}) and (\ref{dilaton}) in conformal gauge takes the form

\be
S_{5D}=\frac{1}{\pi \sdrs} \int d^2z \Bigl[ G_{MN}(X)+B_{MN}(X) \Bigr]\partial
X^M \bar{\partial} X^N +\sdr {\cal R}^{(2)} \phi(X), \ee where $G_{MN}$ is the
$5d$ metric with $M,N=0 \ldots 5$, ${\partial}$ ($\bar{\partial}$) is the left
(right) derivative, and the background dilaton and anti-symmetric tensor are
denoted $\phi$ and $B_{MN}$, respectively. In order to reduce this theory on a
circle of radius $R$, let us consider the following factorizable background
metric
\be
ds^2=G_{MN}=-g^{(4)}_{\mu \nu} dx^\mu dx^\nu +R^2 dy^2. \ee Using this metric
in the above action we find \bea \label{41met} S_{4D+1}&=&\frac{1}{\pi \sdrs}
\int  d^2z \Bigl[ G_{\mu \nu}(X)+B_{\mu \nu}(X) \Bigr]\partial X^{\mu}
\bar{\partial} X^{\nu} +\Bigl[ G_{\mu 5}(X)+B_{\mu 5}(X) \Bigr] \bar{\partial}
X^{\mu} {\partial} X^{5} \nonumber \\ &&+\Bigl[ G_{\mu 5}(X)-B_{\mu 5}(X)
\Bigr] {\partial} X^{\mu} \bar{\partial} X^{5} + G_{5 5}(X)
\partial X^{5} \bar{\partial} X^{5} +\sdr {\cal R}^{(2)} \Phi(X), \eea where $R \equiv \sqrt{G_{55}}$ is the
radius of the extra dimension. The mass of the string state is given as before
from (\ref{lmatch}) and (\ref{tstringenergy}), with $a_L=a_R=-1$ since we are
considering bosonic strings. The mass and level matching are then given by \bea
\label{mass} M^2=\frac{n^2}{R^2}+\frac{\omega^2 R^2}{\alpha^{\prime
2}}+\frac{2}{\alpha^{\prime}}(N_L+N_R-2), \nonumber \\ n\omega+N_L-N_R=0, \eea
where the integers $n$ and $\omega$ label the momentum and winding charge
associated with the extra dimensions and $N_L$ ($N_R$) correspond to the number
of left (right) oscillators that are excited, which can be taken in the compact
$N^{(5)}_L, N^{(5)}_R$ or non-compact directions $N^{(\mu)}_L, N^{(\mu)}_R$.

We are interested in the low-energy or massless states given by (\ref{mass}).
For generic radii no non-trivial winding or momentum is allowed, i.e.
$n=\omega=0$. If the oscillators are restricted to the non-compact dimensions,
i.e. $N^{(5)}_L=N^{(5)}_R=0$, we have the $4D$ graviton, flux, and dilaton.  If
the oscillators are taken in the compact direction we get one scalar (the
radion)
\be
\sigma=\ln \left( \frac{R}{\sdr} \right), \ee and two vectors \bea
A^\mu_{\text{left}} \equiv A^\mu =\frac{1}{2} \left( G_{\mu 5}+B_{\mu 5}
\right),\nonumber \\ A^\mu_{\text{right}} \equiv \bar{A}^\mu=\frac{1}{2} \left(
G_{\mu 5}-B_{\mu 5} \right). \eea To find the evolution of the fields, we
calculating the beta equations for the action (\ref{41met}) and demand that the
couplings do not spoil conformal invariance \cite{Bagger:1997dv}. In the low
energy limit these equations can be derived from the usual space-time action
for dilaton gravity with flux (\ref{theaction}) with an additional contribution
coming from the fields above given by \be \label{abelian} S_m=\int d^4x
\sqrt{G} \left[ (\partial \sigma)^2-\frac{1}{4g^2}(F_{\mu \nu}F^{\mu \nu})
-\frac{1}{4{g}^2}(\bar{F}_{\mu \nu}\bar{F}^{\mu \nu}) \right], \ee where the
abelian field strength is given by $F_{\mu \nu}=\partial_{\mu}
A_\nu-\partial_{\nu} A_\mu$ and $\bar{F}_{\mu \nu}=\partial_{\mu}
\bar{A}_\nu-\partial_{\nu} \bar{A}_\mu$. In addition, the beta equations
naturally enforce the Lorentz gauge condition
\be
\partial_\mu A^{\mu}=0.
\ee Thus, the low energy theory of a bosonic string compactified on
$\mathcal{M}^4 \times S^1$ is described by $4D$ dilaton-gravity with flux
coupled to a chiral $U(1)$ gauge theory.

Now let us consider the mass spectrum at the self-dual radius $\sigma=0$.  In
this case the mass and constraint (\ref{mass}) become \bea \sdrs
M^2=(n+\omega)^2 +4(N_L-1),\nonumber \\ n\omega+N_L-N_R=0, \eea leading to the
additional massless states;
\begin{center}
\begin{tabular}{|l|llllll|}
\hline Scalars&$N_L^{(\mu)}$&$N_R^{(\mu)}$&$N_L^{(5)}$&$N_R^{(5)}$&$n$&$w$\\
\hline &0&0&0&0&0&$\pm 2$\\ &0&0&0&0&$\pm 2$&0\\ &0&0&1&0&$\pm 1$&$\mp 1$\\
&0&0&0&1&$\pm 1$&$\pm 1$\\ \hline \hline
Vectors&$N_L^{(\mu)}$&$N_R^{(\mu)}$&$N_L^{(5)}$&$N_R^{(5)}$&$n$&$w$\\ \hline
&1&0&0&0&$\pm 1$&$\mp 1$\\ &0&1&0&0&$\pm 1$&$\pm 1$\\ \hline
\end{tabular}
\end{center}
These new states combine with the previous scalar and vectors to fill out the
adjoint representation of $SU_L(2) \times SU_R(2)$ \cite{Bagger:1997dv}. Thus,
for arbitrary radius the matter action is given by the chiral $U(1)$ gauge
theory (\ref{abelian}), and as we approach the ESP (self-dual radius) the
theory is lifted to a non-Abelian chiral $SU(2)$ gauge theory. In the latter
case the field strengths are now given by the Yang-Mills theory
\be
F^a_{\mu \nu}=\partial_{\mu} A^a_{\nu}-\partial_{\nu} A^a_{\mu}+
g\epsilon^{abc}A^b_{\mu}A^c_{\nu}, \ee
\be
\bar{F}^a_{\mu \nu}=\partial_{\mu} \bar{A}^a_{\nu}-\partial_{\nu}
\bar{A}^a_{\mu}+ g\epsilon^{abc}\bar{A}^b_{\mu}\bar{A}^c_{\nu}, \ee and the
scalars couple through the ({\bf a},{\bf 0}) and ({\bf 0},{\bf a}) gauge
covariant derivatives \be \label{gcd}
(D_{\mu}\phi)^a=\partial_{\mu}\phi^a+g\epsilon^{abc}A^b_{\mu}\phi^c, \ee \be
\label{gcd2} (\bar{D}_{\mu}\phi)^a=\partial_{\mu}\phi^a
+g\epsilon^{abc}\bar{A}^b_{\mu}\phi^c, \ee where the coupling $g$ is of ${\cal
O}(1)$ for the states we are considering\footnote{For example, for the
heterotic string the four dimensional gauge coupling is given by $g^2=4
\kappa^2 / \sdr$, where $\kappa$ is the gravitational length and contains the
dilaton expectation value.  One can usually choose these values so that $g$ is
order one, which is expected from the Yang-Mills theory.  This implies that the
string scale is close to the gravitation scale.  For a complete discussion see
\cite{Polchinski:1998rq,Polchinski:1998rr}.} and $\phi^a$ is in the ({\bf
3},{\bf 3}) adjoint representation of the chiral $SU(2)$.  The gauged kinetic
term leads to an effective mass for the vectors $m^2_A \sim g^2 \sigma^2$ and
similarly for the additional scalars. Thus, we see that the radion is acting to
give masses to the string states in the same way as the Higgs particle in
ordinary gauge theories with spontaneously broken symmetries
\cite{Polchinski:1998rq}.

It was observed in \cite{Watson:2004aq} that considering this effect for
homogeneous, but time-dependent fields can lead to a stabilization mechanism
for the radion.  For simplicity let us take the dilaton to be fixed and using
the adiabatic approximation, let us consider strings in a $4D$ FRW universe
with metric
\be
ds_4^2=-dt^2+e^{2\lambda(t)}d\vec{x}^2. \ee

The effective action for generic $\sigma$ is given by \be \label{etheaction}
S_{eff}=\int d^4x \sqrt{g}\; \Bigl[ R- \frac{1}{2}(\partial \sigma)^2 - V_{eff}
\Bigr], \ee where $V_{eff}$ initially represents the contribution from the
chiral $U(1)$s, although near the self dual radius it should incorporate the
effects due to the additional massless states.

Let us consider the background equations of motion first, neglecting the
backreaction near the ESP.  The equations following from (\ref{etheaction}) are
\bea \label{eqn1} 3\dot{\lambda}^2&=&\frac{1}{2}\dot{\sigma}^2+\rho_{sub},
\\
2\ddot{\lambda}+3\dot{\lambda}^2&=&-\frac{1}{2}\dot{\sigma}^2-p_{sub},
\\ \label{eqno1}
\ddot{\sigma}+3 \dot{\lambda} \dot{\sigma}& =&\frac{\partial V_{eff}}{\partial
\sigma}, \eea where $\rho_{sub}$ and $p_{sub}$ represent the subdominant
contribution from the $U_L(1) \times U_R(1)$ contained in $V_{eff}$ at generic
radii. This contribution will be subdominant at early-times, since the kinetic
term has an equation of state $\rho=p$ and thus scales as $\rho=a^{-6}$.  The
corresponding scale factor is $a(t) \sim t^{1/3}$ and $\dot{\lambda}=1/3t$.  In
this limit we can ignore the potential in (\ref{eqno1}) and $\sigma$ is given
for small $t$ as
\be
\sigma(t)=\sigma_0+v_0 t. \ee We start the time evolution at $t=0$ when the
field is closest to $R=\sdr$, thus we see that $\sigma_0$ is a measure of how
close the radion comes to the ESP.  In the previous section it was shown that
by including the dilaton in the dynamics, along with the winding and momentum
modes of the string, the radion will naturally pass through $\sigma=0$ and be
localized around the ESP. Motivated by this result we assume $\sigma_0=0$,
which is the most efficient case for particle production, since the states will
be exactly massless there.

We proceed to address particle creation in a way analogous to (p)reheating in
so-called NO (No Oscillation) models of inflation
\cite{Felder:1998vq,Felder:1999pv}.  The method of quantum trapping was first
discussed in \cite{Kofman:2004yc}, where the application of the trapping was
applied to a D-brane moduli space with the trapped modulus corresponding to the
separation of two D-branes and the light states corresponding to open strings
stretched between the branes which become massless as the branes approach.
Since we are discussing the creation of strings, one might wonder if we are
justified in taking the field theoretic approach that is usually utilized in
models of reheating. This issue was addressed in \cite{Gubser:2003vk}, where it
was shown that the effective field theory is adequate to describe string
production mode by mode in a way analogous to the usual point particle case.
Using this approach, we can think of each string mode as a scalar field with a
time varying mass.

For example, let us consider the effects of producing one of the additional
massless vectors that appear at the ESP.  From the coupling in (\ref{gcd}) we
see that the additional states would lead to a potential \be
\label{thepotential} V_{eff}(\sigma,A_{\mu})=\frac{1}{2}(\partial_{\mu}
A_{\nu})^2-\frac{1}{2} g^2 \sigma^2 A_{\mu}A^{\mu}, \ee where $A_{\mu}$ is one
of the additional massless vectors. Note that we are neglecting the other
Yang-Mills interactions, as these would lead to the same generic dynamics for
$\sigma$. However, it would be interesting to include these interactions in
future work, as they are examples originating directly from string theory of
the type of interactions recently considered in \cite{Gubser:2004uh} as dark
matter candidates. We will discuss this possibility in the next section in some
detail.

From (\ref{thepotential}), we can identify $m(t)^2=g^2 \sigma^2$ as a time
dependent mass for $A_{\mu}$. As $\sigma$ approaches the ESP, the $A_{\mu}$'s
become massless and easy to create. Then, as $\sigma$ leaves the ESP these
states will grow massive.  Considering this backreaction results in an
attractive force pulling $\sigma$ back towards the ESP.

Let us consider the time dependent frequency of a particular Fourier mode
$A^{\mu}_k$
\be
\omega_k(t)=\sqrt{\vec{k}^2+g^2 \sigma^2(t)}. \ee A particular mode becomes
excited when the non-abiabaticity parameter satisfies
${\dot{\omega}}/{\omega^2} \geq 1$. When this condition holds for a particular
mode, it results in particle production and an occupation number
\be
n_k=\exp{ \Biggl( -\frac{\pi \vec{k}^2+g^2 \sigma_0^2}{g v_0}} \Biggr). \ee

Recall that we can take $\sigma_0=0$, while $g$ is a positive constant of order
unity in string units. The energy density of produced particles is given by
\be
\rho_{A} =\int \frac{d^3k}{(2\pi)^3}n_k \omega_k \approx g | \sigma(t) | N, \ee
with $N \sim (gv_0)^{3/2}$. Thus, comparing this to (\ref{eqn1}) we see that
the initial kinetic energy associated with the radion $\frac{1}{2} v_0^2$ is
dumped into production of $A_{\mu}$ particles as the radion passes through the
ESP.  Given a large enough $v_0$, the radion will continue its trajectory and
the modes will become massive as we have seen. This results in an always
attractive force of magnitude $gN$ pointing the radion back towards the ESP.
The effective equation for $\sigma$ including the backreaction is then given by
\bea \label{eqn2} \ddot{\sigma}+3 \dot{\lambda} \dot{\sigma} =-g N(t). \eea
This process will continue with each pass of the radion, until all of its
initial kinetic energy has been used up and it settles to the self dual radius.
Therefore, we are led to the conclusion that the additional states associated
with the enhanced symmetry result in a fixed value for the radion at the self
dual radius.

One immediate concern might be whether this method is stable to perturbations.
Moreover, one could worry that the initial kinetic energy of the radion is so
high that the force associated with the backreaction is not enough to over come
its inertia.  Both of these problems are overcome by considering the Hubble
friction associated with the second term in (\ref{eqn2}).  One expects this
friction to damp out any perturbations and to actually enhance the
stabilization mechanism. This was discussed in models of string gas cosmology
\cite{Battefeld:2004xw} and a similar conclusion was reached in
\cite{Kofman:2004yc}.  Moreover, it was shown in \cite{Battefeld:2004xw} that
once we switch to the effective theory the Hubble friction is enough to keep
the radion evolving slowly compared to the growth of the three large
dimensions.  We conclude that Hubble friction combined with the ESP
backreaction should be more than adequate to stabilize the radion at the self
dual radius.

Despite this promising result for stabilizing the radion, the dilaton still
remains a serious challenge.  One approach to stabilizing the dilaton would be
to search for enhanced symmetry states that depend on the value of the dilaton
in much the same way they did for the radion.  However, this is problematic,
since it requires a knowledge of the effective theory for all values of the
string coupling (dilaton).  One way to circumvent this is to search for
additional light BPS states, since such states are non-perturbative in the
sense that they are understood for all values of the coupling.  Preliminary
results suggest that dynamical stabilization of the dilaton may be possible by
considering certain bound states of membranes in M-theory \cite{Cremonini:2006sx}.  These
membranes have a tension that depends on the radius of the $11th$ dimension,
which is related to the dilaton upon compactification to $10D$ string theory.
This suggests that one could stabilize the dilaton at locations where the
membrane tension vanishes in much the same way as the radion above. One
challenge in this case is understanding the production of string states, since
this depends crucially on the string coupling.  Moreover, as the string
coupling changes there can be competing effects governing the dynamics in
moduli space.  It was shown in \cite{Silverstein:2003hf}, that at strong
coupling corrections to moduli trajectories from virtual effects of the ESP
states can have a more important effect than on-shell production.  This could
actually slow the modulus before it finally reaches the ESP.  Thus, we learn
that the dynamics of moduli can be quite rich if we go beyond the usual static
moduli space approximation.  One might hope that with further investigation and
a better understanding of the dynamics of moduli space the need to resort to
anthropic arguments or a landscape might be avoided. Instead, the universe
could be determined through the effect of string dynamics on a time-dependent
background.

\section{Late time cosmology and observations}
So far, our main concern has been the impact of strings and branes on the
evolution of moduli fields, either at the classical or the quantum level. We
have seen the emergence of possible mechanisms to stabilize moduli fields at
various instances. Given a stabilizing mechanism, e.g. provided by a classical
gas of massless string modes or by quantum trapping as outlined in the previous
section, we can turn our attention to late time cosmology \footnote{The
consistency of the stabilization mechanism in the presence of matter was shown
quantitatively in \cite{Ferrer:2005hr}, where it was also noted that it is not
consistent with the presence of a cosmological constant -- however, an
explanation of the currently observed late time acceleration via the dynamics
of the radion seems possible within SGC \cite{Ferrer:2005hr} (see also
\cite{Biswas:2005wy}).}, and search for observational imprints. Two interesting
possibilities naturally surface:
\begin{enumerate}
\item If a string gas is responsible for stabilizing internal dimensions today and it is taken
in the dark sector, this naturally leads to a candidate for cold dark matter.
\item It is widely believed that some period of cosmological inflation must have occurred
in the past and it seems unavoidable to incorporate inflation into SGC.
However, inflation must have taken place before the moduli were stabilized by
the string gas -- otherwise the gas would have been diluted too much to
effectively stabilize the radion. Since the observed large scale structure of
the universe evolved from quantum fluctuations seeded during inflation, it is
possible that the string gas left observable imprints on the spectrum of
fluctuations.
\end{enumerate}
Both avenues are in their initial stages of being examined and a lot of work
needs to be done before an honest prediction can be made. Nevertheless, we will
have a closer look at recent progress in the next two sections.

\subsection{Dark Matter}
Within the framework of SGC we do not expect to observe single strings, because
the model relies on the presence of a gas of strings. Such a gas will appear as
a component of the energy budget of the universe, not as single objects. We
will take the strings to lie in the dark sector, which suggests they may offer
candidates for both dark energy and dark matter. Both dark energy and dark
matter can only be observed via their gravitational interaction \footnote{They
are observable in the spectrum of fluctuations in the CMBR, gravitational
lensing, galaxy rotation curves, etc.}, but they differ in their equation of
state $p=w\rho$: Dark energy has $w$ close to $-1$ (if it is exactly -1, it is
a cosmological constant) and dark matter has either $w=0$ if it is \emph{cold}
(like pressureless dust), or $1/3$ if it is \emph{hot} (like radiation).

In the following, we will provide a general treatment of different cold dark
matter (CDM) types, following closely
\cite{Gubser:2004uh,Gubser:2004du,Nusser:2004qu}. Thereafter, we discuss how
dark matter arises in the framework of SGC \cite{Battefeld:2004xw}, focusing on
a simple realization via a classical gas of winding and momentum modes in a
universe with only one extra dimension.

\subsubsection{General setup}
Our starting point is a low energy effective action, valid at late-times. We
saw in the previous sections how scalars like the radion arise in an effective
four dimensional description, with a potential dictated by the string gas under
consideration. Hence we will focus on a single scalar with action
\begin{eqnarray}
S_{eff}&=& \int d^4x \sqrt{-{g}} \Biggl( \frac{1}{2 l_p^2}  R[{g}_{\mu
\nu}]-\frac{1}{2} {g}^{\mu \nu} {\nabla}_\mu \psi {\nabla}_\nu \psi+
\mathcal{V}(\lambda,\psi) \Biggr)\,,
\end{eqnarray}
where $\mathcal{V}(\psi,\lambda)=\sum_in_im_i$ defines the masses $m_i$ and
number densities $n_i$. We keep the potential and hence the masses $m_i$
general for the time being, and give a concrete example later.

We are interested in the way different dark matter particles interact with each
other and furthermore, how they influence structure formation. With this
knowledge one can then discuss specific imprints onto the large scale structure
of the universe, as was done in \cite{Gubser:2004du,Nusser:2004qu}. The
deviations from standard $\Lambda$CDM models are in the form of an additional
\emph{fifth force}, mediated by the scalar.

From the Klein Gordon equation of motion for $\psi$ one can read off the
magnitude of the force between the dark matter particles of different type
\cite{Gubser:2004uh}
\begin{eqnarray}
F_{ij}=\beta_{ij}\frac{Gm_im_j}{r^2}\,,
\,\,\,\,\,\beta_{ij}=1+\frac{Q_iQ_j}{l_p^2m_im_j}\,,
\end{eqnarray}
where we introduce the scalar \emph{charges}
\begin{eqnarray}
Q_i=\frac{dm_i}{d\psi}\,.
\end{eqnarray}
Since we are considering scalar gravity, we have that like charges attract and
unlike charges repel. We also note that $\psi$ should be stabilized by a
potential $\mathcal{V}$ in order to avoid problems with observations. As a
consequence, charge neutrality
\begin{eqnarray}
\sum Q_in_i=0
\end{eqnarray}
is required. If the charges vanish, as is the case for standard baryonic
matter, we are left with the Newtonian limit of general relativity, as it
should be. It is via its charge that the CDM we are interested in modifies
structure formation.

In order to understand structure formation one first needs to understand how
small initial under- and over-densities grow due to the gravitational
instability. This means we need to study how perturbations in the densities of
each matter type evolve. Since this is not our main focus, we refer the reader
to the literature \cite{Gubser:2004uh} and summarize the main results below.
Let us introduce the density contrast of dark matter
\begin{eqnarray}
\delta_i&=&\frac{\delta \rho_i}{\rho}\,,
\end{eqnarray}
where $\rho_i=n_im_i$ are the densities of dark matter and $\rho$ is the total
density. The equations of motion for the $\delta_i$ become
\begin{eqnarray}
\ddot{\delta}_i+2\dot{\lambda}\dot{\delta}_i=\frac{\rho
l_p^2}{2}\sum_j\beta_{ij}f_j\delta_j\,,\label{eomdensitycontrast}
\end{eqnarray}
where we introduced the mass fraction $f_i=n_im_i/\sum_i n_i m_i$. Once again,
the Newtonian limit is recovered in the case of vanishing charges. We would
like to emphasize that the whole treatment up to this point holds true only for
small curvatures and nonrelativistic dark matter, which is exactly the case we
are interested in at late-times.

By discussing solutions to (\ref{eomdensitycontrast}) one can study how the
large scale structure with its filaments and voids builds up. If one compares
the resulting universe to common $\Lambda$CDM computations and observations,
one has a way of verifying or excluding the existence of a specific string or
brane gas. However, this seems to require improved observations. We conclude
this brief summary and refer the interested reader to
\cite{Gubser:2004du,Nusser:2004qu} where the study of structure formation was
developed in much more detail, and the connection to observations has been
discussed.

\subsubsection{Example: A dark matter candidate within SGC}
We shall now examine a simple example as introduced in \cite{Battefeld:2004xw}.
The goal here is not to present a complete model, but only to suggest how dark
matter may arise from SGC. The generalization to other types of string/brane
gases should be straightforward.

Let us consider the case of only one extra dimension filled with a gas of
winding and momentum modes. Going to the Einstein frame and integrating out the
extra dimension an effective action of type (\ref{effaction4d}) results. Giving
the dilaton a VEV of $\phi=0$, the potential turns out to be
\begin{eqnarray}
\mathcal{V}=\frac{\mu
V_1Ne^{\frac{\sqrt{6}l_p\psi}{6}}}{e^{3\lambda}}+\frac{\mu V_1M
e^{\frac{-\sqrt{6}l_p\psi}{2}}}{e^{3\lambda}}\,,
\end{eqnarray}
after following the procedures of Subsection \ref{section:dilatons}. Here $V_1$
is the spacial volume of the extra dimension (so that $M_p^2=V_1M_5^2$), and
$M$, $N$ are the numbers of winding and momentum modes, respectively. A stable
minimum at the self dual radius $\psi=0$ results, if we have $3M=N$
\footnote{It is then consistent to give the dilaton the VEV we chose.}.  For
other string gases the potential will differ accordingly.

We can now identify the number densities
\begin{eqnarray}
n_1=\frac{M}{e^{3\lambda}}\,, \,\,\,\,\, n_2=\frac{N}{e^{3\lambda}}
\end{eqnarray}
and the masses
\begin{eqnarray}
m_1=\mu V_1e^{-\frac{\sqrt{6}}{2}l_p\psi}\,,\,\,\,\,\,m_{2}=\mu V_1
e^{\frac{\sqrt{6}}{6}l_p\psi}\,.
\end{eqnarray}
The densities scale as $e^{-3\lambda}$, just like matter, so that we can
identify this specific string gas as a CDM candidate. Computing the charges
$Q_i$ one sees that the total charge vanishes at the self dual radius, as it
should. The mass fractions become $f_1=f_2=1/2$ and the $\beta$ matrix is given
by
\begin{eqnarray}
\beta_{ij}= 4 \left[
\begin{array}{rr}
1 & 0 \\ 0 &  \frac{1}{3}
\end{array} \right]\,.
\end{eqnarray}
The matrix is diagonal showing the absence of any long range interaction
between winding and momentum modes. This is consistent with the overall setup.

One can now go ahead and solve the equation of motion
(\ref{eomdensitycontrast}) and follow up with a numerical treatment once the
perturbations become nonlinear. Here we will only mention the modes of
instability in the linear regime \cite{Battefeld:2004xw}; there is an adiabatic
mode and another subdominant mode. The adiabatic mode corresponds to the
movement of strings together with the expansion in the matter dominated epoch.

The addition of other string or brane gases is straightforward, and one can
find rich physics in the dark sector that still needs to be explored in more
detail. Also, a connection to Chameleon cosmology as proposed by Khoury and
Weltman seems possible (see e.g. \cite{Brax:2004px} and references within), but
has not yet been examined.


\subsection{Imprints onto Perturbations}
In this section, we are interested in possible imprints of string gases on
perturbations of the metric degrees of freedom.  These signatures can then be
probed, e.g. via an observation of the cosmic microwave background radiation.

We begin in a phase with three dimensions inflating, while the other
dimensions are deflating. During this phase, metric fluctuations are generated by
the string gases and continue to evolve until the perturbation exits the Hubble radius.
As a consequence, a nearly scale invariant spectrum of fluctuations should result.
Once the internal dimensions evolve to a value where enhanced symmetry occurs, massless string
  modes get produced and these modes can stabilize the internal dimensions as
  we discussed in the last section.  We are then left with a radiation
  dominated FRW universe that is effectively $3+1$ dimensional.
  Then as the universe evolves in the post inflationary epoch, long wavelength
    modes enter the horizon again and leave imprints on the cosmic microwave
    background radiation that we observe today.

The weak point of this proposal is clearly that no successful incorporation of
inflation into the setup of SGC has been realized yet, however efforts in this direction
were considered in \cite{Brandenberger:2003ge,Kaya:2004yj,Parry:2001zg}.
Another possibility is a period of anisotropic inflation as proposed by Levin and others in
 the mid nineties \cite{Levin:1994yw} or more recently in \cite{Patil:2005ii}.
If inflation can be realized, then another
immediate concern arises: Given that a nearly scale invariant spectrum of
fluctuations can be generated during the inflationary phase, one might fear
that the violent production of a string gas at the end of inflation and the
resulting stabilization of the radion will spoil the spectrum.
However, a recent study \cite{Battefeld:2005wv} showed that the spectrum
remains unaltered, which was certainly unexpected. The analysis
was performed in a full 5D setting (the extra dimension being either a circle
or an orbifold), with a classical gas of massless string modes and a radiation
bath present. After finding an approximate analytic solution for the
background, all quantities (the string gas, the radiation bath and the metric)
were perturbed up to first order, the relevant equations of motion derived and
solved (approximate analytical and numerical). The most prominent features of
the solution are the following: long wavelength modes of the Bardeen potentials
(super horizon modes) stay approximately frozen until they re-enter the Hubble
horizon, since the transient oscillations of the radion only source equally
transient oscillations in the Bradeen potentials. The perturbation of the
radion itself exhibits only decaying modes, consistent with a stable radion.
Henceforth, a given spectrum of fluctuations will survive the trapping of the
radion in a similar way as a spectrum survives reheating after standard scalar
field driven inflation.

Based on these results, an important next step within the SGC
program is the incorporation of inflation. This will then allow one to search
for imprints onto the spectrum of perturbation that are unique for SGC.

\section{Summary}
We have seen that an important concept leading to recent progress in SGC is
that of quantum moduli trapping via light states at points of enhanced
symmetry. These states first appeared in SGC while considering the classical
dynamical effects of massive string states containing nontrivial winding and
momentum. The massive states were included in the tree level theory by
including the string sigma model directly in the action to obtain higher order
corrections to the tree level action. This approach was questionable given the
necessary truncation of the string beta equations in order to obtain the low
energy action.  However, it lead to uncovering the importance of additional
massless states that had been missed in the low energy theory.  This is an
example that suggests if we are to build more realistic models of string
cosmology, we really need to go beyond the moduli space approximation and
obtain a better understanding of time-dependent string solutions. Moreover,
even though the focus of SGC has shifted to massless states for moduli
stabilization, the massive modes may still prove vital, especially if the ideas
of Brandenberger and Vafa are to be realized.  We discussed that current
calculations in the low energy theory suggest that the heuristic argument for
dimensionality may not be realized.  Although, it seems that a better
understanding of the non-perturbative aspects of string theory are needed to be
sure.

At late-times, we saw that not only do string gases near ESPs provide moduli
stabilization through trapping, but that string gases also act as an
alternative candidate for cold dark matter. In addition, the framework for
studying signals in the large scale structure of the universe originating from
this dark sector has already been developed by Gubser and Peebles.

Another conclusion of the string gas approach is that it leads naturally to a
string landscape.  This results from the fact that ESPs are quite common in
moduli space and the moduli stabilization can occur at any one of these ESPs.
In fact, ESPs are ubiquitous in any theory with $N=4$ $D=4$ supergravity as a
low energy limit. This makes moduli trapping a common feature on the landscape
of vacua, but it also leaves the question of a definitive vacuum unanswered.
It should also be noted that one
lesson learned from SGC is that our understanding of moduli space dynamics is
in need of further study.  Moduli trapping is only one of many dynamical
effects that one might anticipate on the landscape, and a better understanding
of the dynamics will perhaps lead to a definitive vacuum after-all. Moreover,
in order to obtain realistic phenomenology we are interested in low energy
vacua with at most $N=1$ SUSY and chiral fermions.  Thus, much remains to be
done if we are to build more realistic models, but we hope that we have
demonstrated that SGC offers a framework where many of these questions may be
explored.

\begin{acknowledgments}
We are grateful to Diana Battefeld, Tirthabir Biswas, Robert Brandenberger, Sera
Cremonini, Alan Guth, Liam McAllister, Subodh Patil, and Amanda Weltman for
useful discussions. SW would also like to thank Steve Gubser and Eva
Silverstein for critical comments during the early stages of his work. This
work was financially supported in part by the National Science and Engineering
Research Council of Canada and in part by the U.S. Department of Energy under
Contract DE-FG02-91ER40688, TASK A.
\end{acknowledgments}

\begin{appendix}
\section{Conformal Frames and Dimensional Reduction}
In this appendix we present a brief summary of the methods of dimensional
reduction and conformal transformations. A more complete account can be found
in \cite{Lidsey:1999mc,Birrell:1982ix,Silverstein:2004id,Carroll:2001ih}.  We will use the mostly plus 
convention for the metric $(-+++ \ldots)$ and we follow the
sign conventions of \cite{Wald:1984rg}, denoted (+ + +) in \cite{MTW}.

\subsection{Conformal Transformations}
In general, a conformal transformation
\be
\bar{ds}^2=\Omega^2 ds^2,  \;\;\;\; \bar{g}_{\mu \nu}= \Omega^2 g_{\mu \nu},
\;\;\;\; \bar{g}^{\mu \nu}= \Omega^{-2} g^{\mu \nu}, \;\;\;\;
\sqrt{-\bar{g}}=\Omega^D \sqrt{-g}, \ee does {\bf NOT} leave the action
invariant and results in the following transformations \bea
\bar{\Gamma}^\lambda_{\mu \nu}&=&\frac{1}{2}\Bigl( \bar{g}_{\mu \kappa,
\nu}+\bar{g}_{\nu \kappa, \mu} -\bar{g}_{\mu \nu, \kappa} \Bigr)\\ &=&
\Gamma^\lambda_{\mu \nu}+ \frac{1}{\Omega} \Bigl( g_\mu^\lambda \Omega_{, \nu}
+ g_\nu^\lambda \Omega_{, \mu} - g_{\mu \nu} g^{\lambda \kappa} \Omega_{,
\kappa} \Bigr),\\ \bar{R}&=& \Omega^{-2} \Bigl( R-2(D-1)\Box \ln
\Omega-(D-2)(D-1) g^{\mu \nu} \frac{\Omega_{, \mu}\Omega_{, \nu}}{\Omega^2}
\Bigr),\\ \bar{\Box} \phi &=& \Omega^{-2} \Bigl( \Box \phi+(D-2)g^{\mu \nu}
\frac{\Omega_{, \mu}}{\Omega} \phi_{, \nu} \Bigr), \eea where quantities with a
bar denote the new frame. We can invert to find the old Ricci scalar in terms
of the new one
\be
R=\Omega^2 \Bigl(\bar{R}+2(D-1) \bar{\Box} \ln \Omega-(D-2)(D-1) \bar{g}^{\mu
\nu} \frac{\Omega_{, \mu} \Omega_{, \nu}}{\Omega^2}  \Bigr). \ee We see that if
we begin with an action
\be
S=\int \sqrt{-g} f[\phi(x^\mu)] \Bigl(R+ \ldots \Bigr) \ee for a general
modulus field $f[\phi(x^\mu)]$ multiplying the Ricci scalar, the term can be
transformed to the canonical Einstein frame by choosing
$\Omega^2=f^{\frac{2}{D-2}}$.
\\
\\
{\bf String frame to Einstein frame}\\ \indent As an example, consider starting
with the bosonic string frame action in $D$ dimensions
\be
S_S=\frac{1}{2 \kappa^2}\int d^{D}x \sqrt{-G} e^{-2 \phi} \Bigl( R+4(\partial
\phi_S)^2 -\frac{1}{12}H^2 \Bigr). \ee We can then go to the Einstein frame by
the transformation
\be
g^E_{\mu \nu}=\Omega^2 g^S_{\mu \nu}, \;\;\; \Omega^2=\exp \Bigl( -\frac{4
\phi}{D-2} \Bigr), \;\;\; \phi_E=2\sqrt{\frac{2}{D-2}}\; \phi, \ee with the
field redefinition making $\phi_E$ canonical. The action becomes
\be
S_E=\frac{1}{16 \pi G_D}\int d^{D}x \sqrt{-G} \Bigl( R-\frac{1}{2}(\partial
\phi_E)^2 -\frac{1}{12}e^{-2\sqrt{\frac{2}{D-2}}\phi_E} H^2\Bigr), \ee where
factors of $g_s^2$ and $\alpha^{\prime}$ are present in the $D$ dimensional
Newton constant $G_D$ and $\phi_E$ is the scalar fluctuation associated with
the dynamical dilaton.

\subsection{Dimensional Reduction}
Consider the toriodal compactification of the bosonic degrees of freedom with
action \be \label{appaction} S_{D+d}=\frac{1}{2 \kappa_{D+d}^2}\int d^{D+d}x
\sqrt{-G_{D+d}} e^{-2\phi} \Bigl( R_{D+d}+4(\partial \phi)^2-\frac{1}{12}H^2
\Bigr), \ee where $G_{D+d}$ is the higher dimensional metric, $\phi$ is the
dilaton, and $H=dB$ is the NS three form field strength of the fundamental
string. For this toriodal compactification the geometry is factorizable
$\mathcal{M}_{D+d}=\mathcal{M}_D \times \mathcal{T}_d$ with metric
\be
ds_{D+d}^2=g_{\mu \nu} dx^{\mu} dx^\nu + h_{ab} dy^a dy^b, \ee where $g_{\mu
\nu}$ is the metric on $\mathcal{M}_D$ parameterized by coordinates $x^\mu$,
and $h_{ab}$ is the metric on the compactified space $\mathcal{T}_d$ with
periodic coordinates $y^a$. We will assume that all matter fields are at most
functions of the $x^\mu$, e.g. $\phi=\phi(x^\mu)$. This implies that the
compact space must be Ricci flat, and we will further assume the flux $B$ is
block diagonal. Given this metric, the Ricci scalar will factorize as \be
\label{Ddricci} R_{D+d} = R_D +\frac{1}{4} \nabla_{\mu} h^{ab} \nabla^{\mu}
h_{ab} + \nabla_{\mu} (\ln \sqrt{h}) \nabla^{\mu} (\ln \sqrt{h} )
-\frac{2}{\sqrt{h}} \Box \sqrt{h}, \ee where we used the relation
$\partial_{\mu} \ln h =h^{ab}
\partial_{\mu} h_{ab}$. Plugging (\ref{Ddricci}) into the action (\ref{appaction}) and defining the lower
dimensional dilaton
\be
\varphi \equiv 2\phi-\frac{1}{2} \ln \det h_{ab}. \ee we find \bea
\label{lowerdimaction} S_D=\frac{1}{2 \kappa_D}\int d^D x \sqrt{-g_D}
e^{-\varphi} \Biggl[ R_D +\nabla_\mu \varphi \nabla^\mu \varphi
-\frac{1}{12}H_{\mu \nu \lambda}H^{\mu \nu \lambda} +\frac{1}{4} \nabla_\mu
h^{ab} \nabla^\mu h_{ab} \nonumber \\ -\frac{1}{4}\nabla_\mu B^{ab}\nabla^\mu
B_{ab} \Biggr], \eea where $2 \kappa_{D}^2=2 \kappa_{D+d}^2
\mathcal{V}_0^{-1}=2 \kappa_{D+d}^2 (2\pi \sdr)^{-d}$ and we have defined the
$d$ dimensional volume as
\be
{V}_d={V}_0\int{d^dy} \sqrt{\det h_{ab}} \; ={V}_0 h^{1/2}(x^\mu), \ee where we
used the fact that $h(x^\mu)$ does not depend on the $y^a$ and its components
will appear, along with the $B_{ab}$, as fluctuating scalars in the $D$
dimensional theory. The constant ${V}_0$ is a reference volume and for a string
scale compactification given by ${V}_0=(2 \pi \sdr)^d$. The lower dimensional
Newton constant is then given by
\be
\frac{1}{16 \pi G_{D}}=\frac{{V}_0}{16 \pi G_{D+d}}. \ee

We would like to put (\ref{lowerdimaction}) in Einstein canonical form, which
is accomplished by the conformal transformation \be \label{stoe}
\tilde{g}_{\mu\nu} = \Omega^2 g_{\mu\nu} , \qquad \Omega^2 \equiv \exp \left[ -
\frac{2}{D-2} \varphi \right], \ee and a field redefinition canonically
normalizes the lower dimensional dilaton \be \label{normdilaton} \tilde\varphi
\equiv \sqrt{\frac{2}{D-2}} \varphi \ee which gives the desired form \bea
\label{geq} S=\frac{1}{16 \pi G_D}\int d^D x\sqrt{-\tilde{g_D}} \left[
\tilde{R}_D -\frac{1}{2} \left( \tilde{\nabla} \tilde\varphi \right)^2
-\frac{1}{12} e^{-\sqrt{8/(D-2)}\tilde\varphi} \tilde{H}_{\mu\nu\lambda}
\tilde{H}^{\mu\nu\lambda} \right. \nonumber \\ \left. +\frac{1}{4}
\tilde{\nabla}_{\mu} h_{ab} \tilde{\nabla}^{\mu} h^{ab} -\frac{1}{4}
\tilde{\nabla}_{\mu}B_{ab}\tilde{\nabla}^{\mu} B_{cd} h^{ac}h^{bd} \right],
\eea where we have used $2 \kappa^2_D=16 \pi G_D$.

Now let us specialize this result to the case of an isotropic internal metric,
where the radion is the only degree of freedom.  In this review we have
primarily been interested in the case of vanishing flux ($H=0$) and we started
with the string frame metric
\be
ds^2=-dt^2+a(t)^2 d^2x+b^2(t) dy^2, \ee where $b(t)$ is the $10D$ string frame
radion. By noting $h_{ab}=b^2$ and plugging this result into (\ref{geq}), along
with $D=4$ and neglecting flux, we find
\be
S=\frac{1}{16 \pi G}\int d^4 x\sqrt{-{g}} \left[ {R} -\frac{1}{2} \left(
{\partial} \varphi \right)^2 -d b^{-2}({\partial} b)^2 \right]. \ee We can
canonically normalize the radion by the field redefinition \be \label{candil}
\psi=\sqrt{2d} \ln b, \ee so that we arrive at the desired action
\be
S=\frac{1}{16 \pi G}\int d^4 x\sqrt{-{g}} \left[ {R} -\frac{1}{2} \left(
{\partial} \varphi \right)^2 -\frac{1}{2}({\partial} \psi)^2 \right], \ee where
the four dimensional dilaton is given by \be \label{shiftdil}
\varphi=2\phi-\sqrt{\frac{d}{2}}\psi. \ee Finally, we would like to consider
the addition of a potential term allowing for the presence of strings, branes,
or other matter. If we begin with the potential in the string frame,
\be
S_m^{(4+d)}=-\int d^{4+d}x \sqrt{G_{4+d}} \; \mathcal{V}_s^{(4+d)}, \ee after
the reduction we have \be \label{g4} S_m^{(4)}=-(2\pi \sdr)^d \int d^{4}x
\sqrt{g_{4}} \; b^d \; \mathcal{V}_s^{(4+d)}. \ee Now performing the
transformation (\ref{stoe}) to convert to the Einstein frame the action becomes
\be
S_m^{E}= -\int d^{4}x \sqrt{\tilde{g}_{4}} \; e^{2 \varphi} b^{d}\;
\tilde{\mathcal{V}}_s^{(4+d)}, \ee where we note that the transformation
(\ref{normdilaton}) is trivial in four dimensions, i.e.
$\tilde{\varphi}=\varphi$, and we have absorbed the constant prefactor in
(\ref{g4}) into the potential. To illustrate the scaling with volume and
coupling, let us restore the unshifted dilaton and compact volume using
(\ref{shiftdil}) and (\ref{candil})
\be
S_m^{E}=-\int d^{4}x \sqrt{\tilde{g}_{4}} \; \frac{e^{4 \phi}}{b^{2d}} \; b^d
\; \mathcal{V}_s^{(4+d)}. \ee The final reduced action in the Einstein frame is
\be
S=\int d^4 x\sqrt{-{g}} \left[ \frac{1}{16 \pi G}\Bigl( {R} -\frac{1}{2} \left(
{\partial} \varphi \right)^2 -\frac{1}{2}({\partial} \psi)^2 \Bigr)- e^{4 \phi}
\; e^{-\sqrt{\frac{d}{2}}\psi} \; \mathcal{V}_s^{(4+d)}\right]. \ee Thus, we see
the potential is diluted as the volume runs to large values or the dilaton runs
to weak coupling.  Unfortunately, it is in these limits that string cosmology
is best understood and string corrections are understood. Moreover, if the
potential $\mathcal{V}_s^{(4+d)}$ does not contain large enough powers to
overcome the dilaton and radion, then a local minimum for stabilization is not
found.
\end{appendix}


\bibliography{sgc}

\end{document}